\documentclass[letterpaper]{aa}
\usepackage{txfonts,natbib}
\usepackage{dcolumn}
\usepackage{placeins}
\usepackage[xdvi]{graphicx}
\bibpunct{(}{)}{;}{a}{}{,} 
\newcolumntype{.}{D{.}{.}{4}}
\newcolumntype{,}{D{\cdot}{.}{4}}

\title{Molecular gas in the galaxy cluster Abell~262}
\subtitle{CO observations of UGC~1347 and other galaxies of the cluster} 
\author{T. Bertram \inst{1}
  \and A. Eckart \inst{1}
  \and M. Krips \inst{1}
  \and J.G. Staguhn \inst{2}
  \and W. Hackenberg \inst{3}}
\institute{I. Physikalisches Institut, Universit\"at zu K\"oln,
Z\"ulpicher Str. 77, 50937 K\"oln, Germany 
  \and NASA/Goddard Space Flight Center, Greenbelt, MD 20771, USA
  \and European Southern Observatory, Karl-Schwarzschild-Str. 2, 85748 Garching, Germany}
\offprints{T. Bertram (bertram@ph1.uni-koeln.de)}
\date{Received / Accepted}
\date{Received 17. December 2004 / Accepted 23. September 2005}
\abstract{
We present millimeter CO line emission observations of 12 galaxies within the \object{Abell~262} cluster, together with L$_{\rm FIR}$ data, in the context of a possible molecular gas deficiency within the  region of the cluster center. 
\
Several indications of the presence of such a deficiency are highlighted and connected to a model of cirrus-like cloud stripping.
\
The model predicts a drop in the average 100$\mu$m flux density of galaxies in the core of the cluster compared to the average 100$\mu$m flux density in the outer regions, which is actually indicated in the IRAS data of the cluster members.
\
This drop is explained by the decrease in the total hydrogen column density N(H) and, therefore, also includes a decrease in the molecular gas content.

In addition to results for the global CO content of the galaxy sample, high-resolution interferometric CO(1-0) observations of one of the cluster members, UGC~1347, exemplify the spatial distribution of the molecular gas in a galaxy of the cluster. 
\
With these observations, it was possible to confirm the existence of a bright off-nuclear CO-emission source and to derive molecular masses and line ratios for this source and the  nucleus.
\    
\
}

\begin{document}
\maketitle
\section{Introduction}

A large number of mechanisms in rich galaxy clusters have been discussed \citep{1978apj...226..559b,1998apj...497..188c}, and found to favor the transition from spiral-dominated, blue galaxy populations at higher redshifts towards the domination of early, red populations as observed at z$\sim$0.  
\
The most detailed view of galaxy cluster members has been obtained for the Coma and Virgo galaxies.
\
Studies of these and other targets suggest interaction with either other cluster members or the cluster potential or the hot intra-cluster medium (ICM) seen as the main cause of both the truncation of the star-formation rate and the morphological transition.   
\
Strong evidence of interaction between the interstellar medium (ISM) and the ICM is shown by the lack of atomic gas \citep{1985apj...292..404g} in those spiral galaxies that are in the vicinity of the centers of rich clusters with respect to similar, but isolated objects.
\
This HI deficiency is commonly explained by ram pressure stripping or sweeping, as the galaxy moves through the dense and hot gas in the core of the cluster.      
\
Studies of the global CO distribution of cluster galaxies \citep{1989apj...344..171k, 1991a&a...249..359c, 1998a&a...331..451c, 1997a&a...327..522b,  1998aj....115..405l}, on the other hand, show either no or insignificant dependence of the molecular gas content on its location or on the degree of HI deficiency. 
\
Molecular clouds do not seem, or at least seem less, affected by the gas stripping mechanisms. 
\
The higher column densities and the concentration in the center of the disks and, therefore, the stronger binding  within the galaxies' potential may allow molecular gas to better resist environmental influences.  

Despite this negative global result, CO observations with high spatial resolution can provide important clues to the mechanisms that may influence the evolution of galaxies within a cluster.   
\
\citet{2001a&a...374..824v}, for example, discuss the indirect consequences of HI stripping by studying the lopsided molecular gas distribution in the case of the Coma galaxy NGC~4848.  
\
Although ram pressure stripping, in the long term, leads to an overall decrease in the gas mass, and hence to a reduced star-formation rate (SFR), it can cause a temporary increase in the gas density and, along with this, an increased SFR.    
\
Molecular gas can be regarded as required for the star-formation process.
\
Since CO is the most important tracer for molecular gas, CO observations provide information on whether the environment within cluster members favors enhanced star formation activity or not.

In this paper we present the results of our CO observations of a small sample  of  \object{Abell 262} cluster galaxies and discuss indications for environmental effects on the total molecular gas content of the sample members.
\
With L$_{\rm{FIR}}$ data, we substantiate our tentative findings indicating a dependency of the global molecular gas mass on the distance to the cluster core.
\
For a more profound analysis of the total molecular gas content of Abell~262 galaxies, it is essential to determine the extent and distribution of the CO emission regions with high spatial resolution interferometric observations. 
\
We present the results of such interferometric CO observations of the \object{Abell~262} cluster member \object{UGC 1347} as an example of a more detailed analysis of the galaxy's CO distribution.         
 
\begin{figure}
  \resizebox{\hsize}{!}{\includegraphics{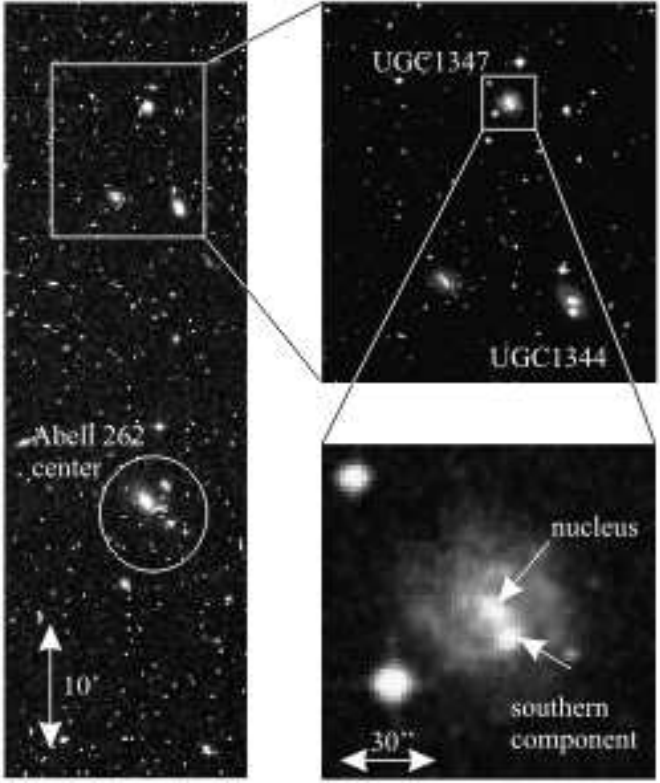}}
  \caption{The location of \object{UGC~1347} within the Abell 262
    cluster -- visual images from the Palomar Observatory Sky
    Survey (POSS). The bottom right image shows the nucleus and the bright southern source at the tip of the bar.}
  \label{abell262-ugc1347}
\end{figure}
\subsection{Abell~262}
The spiral rich cluster Abell 262 has been subject to a variety of studies in
different wavelength domains.
\
\citet{1958apjs....3..211a} first analysed POSS photographic plates, and defined  cluster richness classes by the number of galaxies in the cluster that are not more than 2 magnitudes fainter than the third brightest member. 
\
With 40 galaxies (\citealt*{1989apjs...70....1a}, corrected for background contamination) meeting this richness criterion, Abell~262 falls into richness class 0, and is far less densely populated then e.g. the Coma cluster with 106 members and richness class 2.
\
However, the cluster consists of many more members. \citet{1999apjs..125...35s} consider a total number of 151 galaxies that are associated with Abell~262.
\
As it is embedded in the main ridge of the Pisces-Perseus supercluster \citep{1986apj...306l..55h}, Abell~262 has a conspicuous appearance. 
\
Its center [R.A. (J2000)~=~01$^\mathrm{h}$~52$^\mathrm{m}$~50$^\mathrm{s}$, Dec.
(J2000)~=~36\degr~08.8\arcmin] coincides with the position of the cD galaxy NGC
708. \footnote{Based on ROSAT PSPC observations, \citet*{1996ApJ...473..692D} argue that the position of the central galaxy deviates slightly from center of the cluster potential.}

The value for the cluster redshift of z=0.0163 \citep{1999apjs..125...35s} results in a heliocentric velocity of  v=c$\cdot$z= 4887 km s$^{-1}$ and a distance to the cluster of 49 $h^{-1}$ Mpc. An angular separation of 1\arcsec~
then corresponds to 238 $h^{-1}$ pc. 
\
The Abell radius (corresponding to $\sim$1.5 $h^{-1}$ Mpc) follows r$_A$=1.7\arcmin
$\cdot$ z$^{-1} \simeq$ 1.75\degr.
\
In these and the following calculations, whenever required, H$_0$ = 100~$h$~km~s$^{-1}$~Mpc$^{-1}$ and q$_0$=0.5 were used, unless otherwise stated. 
\
\citet{1996ApJ...473..670F} state a line-of-sight velocity dispersion for
Abell 262 members of $\sigma$ = 525$^{+47}_{-33}$ km s$^{-1}$, which
is similar to the result of \citet{2001ApJ...548..550N} ($\sigma$ =
548 $\pm$ 36  km s$^{-1}$ for a sample of 101 Abell 262 member
galaxies) and is only slightly higher than the average low-velocity dispersion
($\sigma$ = 485 $\pm$ 45  km s$^{-1}$) of clusters in the richness class 0 \citep{1996ApJ...473..670F}.

Several groups have carried out X-ray observations to investigate the properties of the hot ICM of Abell 262 \citep[e.g.][]{2001ApJ...548..550N,
1996ApJ...473..692D,1996ApJ...468...86M}. \citet{1984ApJ...276...38J}
identified Abell~262 as a dynamically young XD cluster, similar to Virgo,
that shows the typical signs of a less evolved system with a stationary, 
dominant galaxy at its center: low X-ray luminosity 
(3.07 $\cdot$10$^{43}$ ergs s$^{-1}$), low ICM temperature (2.8 $\times$ 10$^7$ K), high central gas density/cooling flow, high spiral fraction, low central galaxy density, and an irregular shape.

Since Abell 262 apparently is not yet in a relaxed state, the transformation mechanisms that are discussed in the context of higher redshift galaxy and cluster evolution may still be active. 
\
And due to the cluster's low redshift, they can be studied with high spatial resolution. 
\
Several studies of the signatures of galaxy interaction have included the case of Abell~262. 
\
Similar to clusters such as Virgo, Coma, and several others, Abell 262
member galaxies are deficient in atomic gas towards the center
(\citealt*{1982apj...262..442g}; \citealt{2001apj...548...97s}).   
\
\citet{1997A&AS..126..537B} used the Westerborg Synthesis Radio Telescope to
map the spatial HI distribution of 11 Abell~262 members. 
\
They show that, in a few cases, the gas distribution is asymmetrical -- again a sign of ISM-ICM interaction. 
\
A few Abell~262 galaxies were included in CO(1-0) observations by \citet{1998aj....115..405l}, and H$\alpha$ emission of several members was investigated by \citet{1994a&as..103....5a} and \citet{2000mnras.317..667m}.
\
\citet{2000mnras.317..667m} discuss the H$\alpha$ emission of Abell 262
member galaxies together with other cluster galaxies in the context of tidally
induced star formation, whereas \citet{1994a&as..103....5a} provide maps, as
well as rotation curves, of six Abell 262 galaxies -- among them UGC 1347. 
\
\subsection{UGC~1347}
The SBc galaxy \object{UGC~1347} [R.A.
(J2000)~=~01$^\mathrm{h}$~52$^\mathrm{m}$~45.9$^\mathrm{s}$,    
Dec. (J2000)~=~36\degr~37\arcmin~09\arcsec], is located about 28\arcmin~north
of the cluster center (cf. Fig. \ref{abell262-ugc1347}).  
\
Based on HI observations, \citet{1997AJ....113.1197H}
found the galaxy's redshift to be z=0.01849, corresponding to a systemic velocity of 5543 km s$^{-1}$.  
\
The difference between the cluster velocity of 4887 km s$^{-1}$ and the
systemic velocity of UGC~1347 of about 650 km s$^{-1}$ ($\simeq
1.2$ times the cluster's velocity dispersion)  can be explained as motion of the galaxy within
the cluster.

\object{UGC 1347} is the first extragalactic target that was observed
with a laser-guide-star assisted adaptive optics system. 
\
\citet{2000a&a...363...41h} used the MPIA-MPE ALFA AO system at the Calar
Alto 3.5 m telescope to obtain NIR images at subarcsecond resolution.
They discuss their results in the context of  published radio, FIR, and
H$\alpha$ data. 
\
Since millimetric CO observations had not been carried out for this object by
that time, quantities related to the molecular gas content and
distribution had to be estimated.

Besides a bright nuclear source, \object{UGC~1347} features a second prominent
 component about 9\arcsec~ south of the center at the tip of the bar (cf. Fig. \ref{abell262-ugc1347}).  
\
\citet{2000a&a...363...41h} found this region to be compact in the NIR and 
identified it with a region of recent active star formation in the disk. 
\
\section{Observations and data reduction}
For several members of the \object{Abell 262} cluster, we obtained millimeter CO-emission line data for different transitions and, in the case of \object{UGC 1347}, also for different isotopomeres. The observations were carried out with  BIMA and the IRAM 30m telescope

\subsection{BIMA data}
\element[][12]{CO}(1-0) observations  of \object{UGC~1347} were
carried out with BIMA\footnote{Based on observations carried out with the Berkeley-Illinois-Maryland Association (BIMA) observatory, which is supported by NSF grant AST 99-81289.} at Hat Creek, CA, USA in July 2000. 
\
The local millimeter-wave radio interferometer, consisting of 10
antennae with a diameter of 6.1m each, was used in C configuration.
\
Every 24 minutes the series of observations with this setup was interrupted  by
measurement of the  phase calibrator \object{0136+478} located 11.5\degr~
north of \object{UGC~1347}.
\
The good uv coverage obtained in the overall on-source integration time of 257
minutes resulted in an approximate beam size of 6.6\arcsec~ FWHM.  
\

\begin{table}
  \caption[]{BIMA observation parameters}
  \begin{tabular*}{\linewidth}{@{\extracolsep\fill}lllll}
     \hline
     \hline
    obs. &  line                 & spectral      & array         & $\theta_{\rm{Beam}}$ \\
   date  &                       & resolution    & configuration & FWHM   \\
         &                       & [km s$^{-1}$] &               & [\arcsec]\\
      \hline
 Jul. 00 &\element[][12]{CO}(1-0) &  8.3          & C             & 6.6\\
      \hline
  \end{tabular*}
\end{table}

\begin{figure}
  \resizebox{\hsize}{!}{\includegraphics{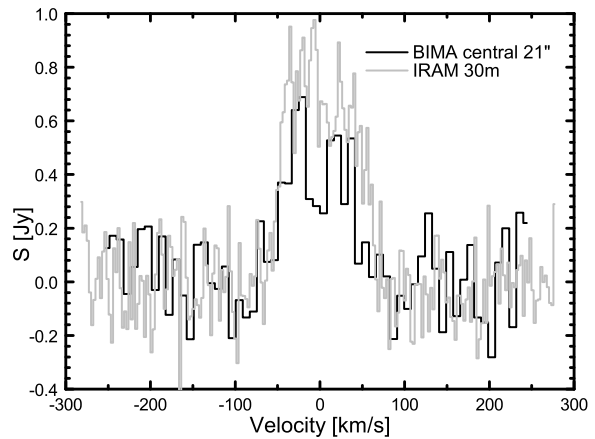}} 
  \caption{Comparison of two \element[][12]{CO}(1-0) spectra, obtained
   with BIMA and with the IRAM 30m telescope. The BIMA flux density
   was integrated over the inner 21\arcsec$^2$, the region covered by
   the 21\arcsec~ diameter IRAM beam. A beam-filling factor $F_{\rm{gauss}}=1+\left(\theta_{\rm{Beam}}/\theta_{\rm{Source}}\right)^2$ was applied to the IRAM 30m data, assuming both beam and source distribution to be Gaussian.} 
  \label{12CO10-spec-IRAM-BIMA}
\end{figure}

Because of the data quality, it was possible to apply all standard reduction steps in an almost straightforward manner.
\
A CLEAN algorithm was applied to the image data obtained.
\
A comparison with single dish data  is shown in  Fig. \ref{12CO10-spec-IRAM-BIMA}.
\
Only about 70\% of the total flux is expected to be recoverable from observations of sources with small angular extensions, such as UGC~1347 in BIMA's  C configuration \citep{2002pasp..114..350h}.

The following remarks on smaller problems in the dataset may be helpful for archive users.  
\
An inspection of the uv dataset revealed that the first channel of each
of the 16 spectral windows was malfunctioning.
\
Both channels neighboring a malfunctioning channel showed reduced dynamics with respect to the other channels. 
\
While the spectral information of the malfunctioning channels was
completely corrupted and therefore lost, the data of the neighboring
channels with damped performance was recoverable by applying a
correction factor. 
\
One of the malfunctioning channels unfortunately represents the  8.3
km/s wide spectral region starting at -4 km/s -- right in the
center of the emission line region of \object{UGC~1347}. 
\
In order to obtain a reasonably integrated intensity distribution
of the galaxy and the integrated flux of the central source, the
missing channel map was interpolated. 
\

\subsection{IRAM 30m data}
\
\begin{figure*}
  \centering
  \includegraphics[height=23.0cm]{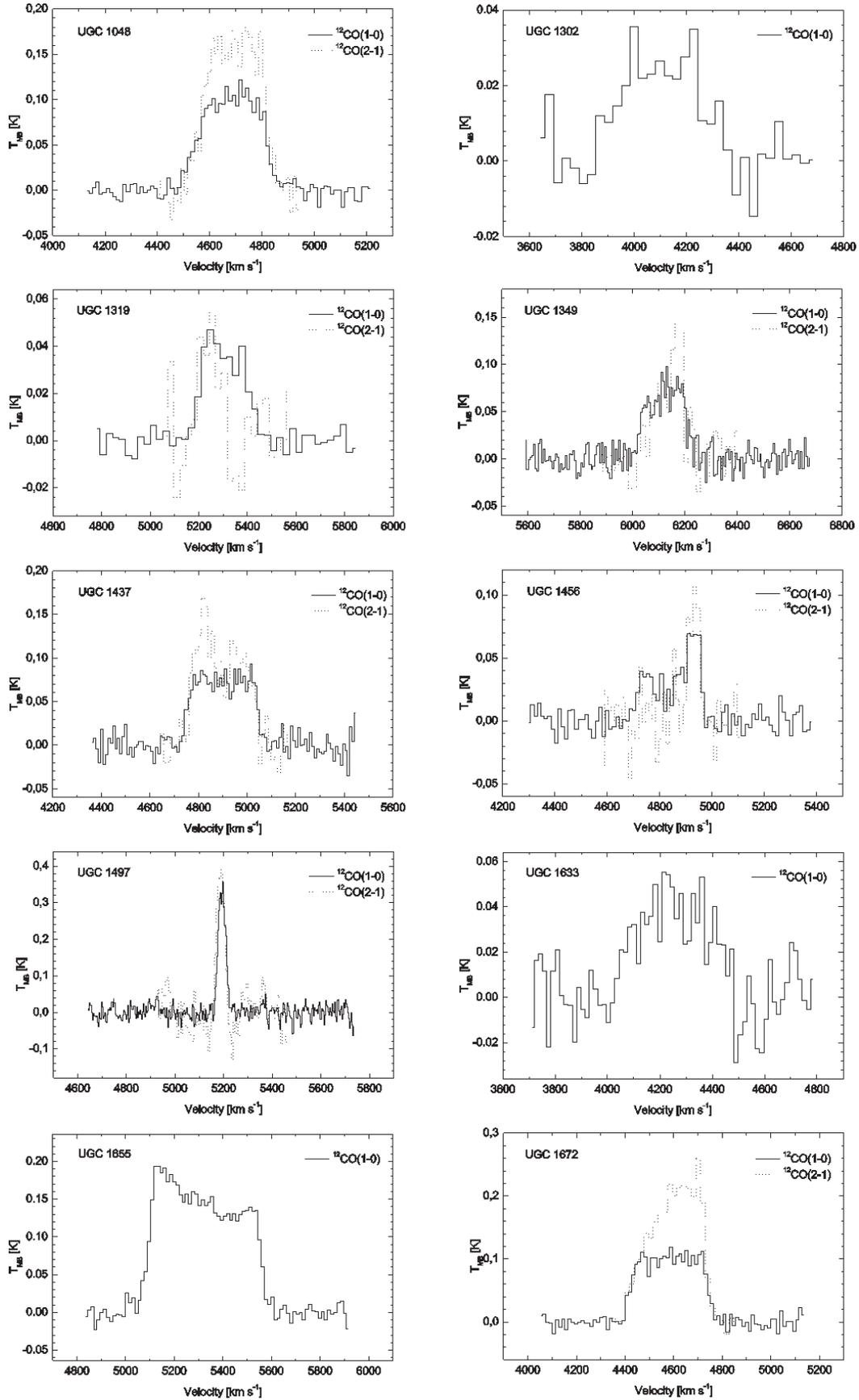}
  \caption{IRAM 30m \element[][12]{CO} spectra of the selected Abell 262 galaxies. The cases in which the \element[][12]{CO}(2-1) emission line region extended over
  the bandwidth of the receiver have been excluded. In these cases a continuum
  subtraction was impossible.}   
  \label{Center-Spectra}
\end{figure*}
\
The IRAM\footnote{IRAM is supported by the INSU/CNRS (France), the MPG (Germany) and the IGN (Spain)} 30m telescope on Pico Veleta (Spain) was used to acquire
\element[][12]{CO}(2-1) and \element[][13]{CO}(1-0) emission line data
of \object{UGC~1347} in parallel. These observations were carried out in July
2003, together with \element[][12]{CO}(1-0) and
\element[][12]{CO}(2-1) measurements of 11 other Abell 262 cluster 
galaxies.  
\
Besides the observations mentioned, initial IRAM 30m measurements of the \element[][12]{CO}(1-0) and \element[][12]{CO}(2-1) emission of
\object{UGC~1347}  were performed in June 2000, prior to the BIMA
observations.

Except for a few cases in which the line emission extended over the
spectral range of the receiver, linear baselines were subtracted from the spectra (cf. Fig. \ref{Center-Spectra}) and peak temperatures, as well as line intensities, were determined.
\
In order to obtain main beam temperatures, the main beam efficiencies
B$_{\rm{eff}}$ and the forward efficiencies F$_{\rm{eff}}$ given in
Table \ref{IRAM-observation-parameters} were applied to T$_A^*$.
\
The resulting values for I$_{\rm{CO}}$ and peak T$_{\rm{MB}}$, as well as the
velocity ranges over which the lines were integrated, are shown in Table 
\ref{CO-results}. 
\
The errors of T$_{\rm{MB}}$ in Table \ref{CO-results} represent the RMS
per channel after smoothing the spectra.

To compare the two \element[][12]{CO}(1-0) spectra obtained with
BIMA and the IRAM 30m telescope shown in Fig. \ref{12CO10-spec-IRAM-BIMA}, the IRAM 30m telescope T$_{\rm{MB}}$ spectrum was converted using the factor S/T$_{\rm{MB}}$ = 4.95 Jy K$^{-1}$. 
\
The inverse conversion of integrated flux densities measured by BIMA into I$_{\rm{CO}}$, as given in Table \ref{CO-content}, was done analogously with a factor S/T$_{\rm{MB}}$ = 0.46 Jy~K$^{-1}$.
\
\begin{table}
  \caption[]{IRAM observation parameters}
  \begin{tabular*}{\linewidth}{@{\extracolsep\fill}llllllll}
     \hline
     \hline
 obs. &   line                    &rest      &  $ \theta_{\rm{Beam}} $        & B$_{\rm{eff}}$  & F$_{\rm{eff}}$\\
 date &                           &freq.     &   FWHM     &                 & \\
      &                           &$[$GHz$]$ &   [\arcsec]& \\
      \hline
 Jun. 00 & \element[][12]{CO}(1-0) & 115       &  21.5    & 0.80 & 0.93\\
 Jul. 03 &\element[][12]{CO}(2-1) & 230        &  10.5    & 0.52 & 0.91\\
 Jul. 03 &\element[][12]{CO}(1-0) & 115        &  21.5    & 0.74 & 0.95\\ 
 Jul. 03 &\element[][13]{CO}(1-0) & 110        &  22.0    & 0.75 & 0.95\\ 
 \hline
 \end{tabular*} 
 \label{IRAM-observation-parameters}
\end{table}

\begin{table*}
  \caption[]{Measured \element[][12]{CO} properties of the centers of selected
  Abell 262 member galaxies. The corresponding observation parameters are shown
  in Table \ref{IRAM-observation-parameters}.} 
  \begin{tabular*}{\linewidth}{@{\extracolsep\fill}llcrrcrr}
   \hline
   \hline
  & &\multicolumn{3}{c}{ \element[][12]{CO}(1-0)} &  \multicolumn{3}{c}{ \element[][12]{CO}(2-1)} \\
 \multicolumn{1}{c}{Obj.} & \multicolumn{1}{c}{Vel. range} & \multicolumn{1}{c}{spectral Res.} & \multicolumn{1}{c}{peak T$_{\rm{MB}}$}& \multicolumn{1}{c}{I$_{\rm{CO}}$ } & \multicolumn{1}{c}{spectral Res.} & \multicolumn{1}{c}{peak T$_{\rm{MB}}$}& \multicolumn{1}{c}{I$_{\rm{CO}}$ } \\ 
                          & \multicolumn{1}{c}{[km]}       & \multicolumn{1}{c}{[km s$^{-1}$]}    & \multicolumn{1}{c}{ [K]} & \multicolumn{1}{c}{[K km s$^{-1}$]}             & \multicolumn{1}{c}{[km s$^{-1}$]}    & \multicolumn{1}{c}{ [K]} & \multicolumn{1}{c}{[K km s$^{-1}$]}  \\
      \hline
  \object{UGC 1048} &    4450-4850 & 3.25 & 0.10 $\pm$ 0.01  & 28.20 $\pm$ 0.42 & 1.63 &  0.16 $\pm$ 0.04  & 40.72 $\pm$ 0.94 \\
  \object{UGC 1302} &    3850-4400 & 3.25 & 0.03 $\pm$ 0.01  &  9.68 $\pm$ 0.76 & 1.63 & \\
  \object{UGC 1319} &    5150-5500 & 3.25 & 0.04 $\pm$ 0.01  &  8.43 $\pm$ 0.35 & 1.63 &  0.04 $\pm$ 0.02  &  5.02 $\pm$ 1.21 \\
  \object{UGC 1344} &    v$_0$=4170& 3.25 & $<$ 0.01         &                  & 1.63 & $<$0.03 \\
  \object{UGC 1347} &    5450-5650 & 2.6  & 0.07 $\pm$ 0.01  &  4.56 $\pm$ 0.29 & 1.63 &  0.19 $\pm$ 0.04  & 11.22 $\pm$ 1.26 \\
  \object{UGC 1349} &    6000-6250 & 3.25 & 0.08 $\pm$ 0.01  & 12.89 $\pm$ 0.37 & 1.63 &  0.14 $\pm$ 0.02  & 13.76 $\pm$ 1.34 \\
  \object{UGC 1437} &    4720-5060 & 3.25 & 0.08 $\pm$ 0.02  & 22.13 $\pm$ 0.55 & 1.63 &  0.14 $\pm$ 0.01  & 28.47 $\pm$ 1.03 \\
  \object{UGC 1456} &    4650-5000 & 3.25 & 0.06 $\pm$ 0.01  &  9.77 $\pm$ 0.48 & 1.63 &  0.07 $\pm$ 0.01  &  5.67 $\pm$ 0.58 \\
  \object{UGC 1497} &    5150-5250 & 3.25 & 0.32 $\pm$ 0.02  & 12.57 $\pm$ 0.35 & 1.63 &  0.37 $\pm$ 0.05  & 12.78 $\pm$ 1.03 \\
  \object{UGC 1633} &    4000-4500 & 3.25 & 0.04 $\pm$ 0.01  & 13.63 $\pm$ 0.83 & 1.63 &\\
  \object{UGC 1655} &    5050-5600 & 3.25 & 0.19 $\pm$ 0.02  & 71.13 $\pm$ 0.65 & 1.63 &\\
  \object{UGC 1672} &    4400-4800 & 3.25 & 0.11 $\pm$ 0.01  & 32.92 $\pm$ 0.51 & 1.63 &  0.28 $\pm$ 0.05  & 56.40 $\pm$ 1.29 \\
 \hline
  \end{tabular*}
 \label{CO-results}     
\end{table*}

\section{Results}
\subsection{Global CO and L$_{\rm{FIR}}$ properties of Abell~262 galaxies}
\label{SecGlobalCO}
The primary goal of the observations with the IRAM 30m telescope was to
determine the global CO content of a set of gas rich galaxies in the Abell~262
cluster and to investigate a possible dependency on the distance to the cluster
center. 
\
The CO bright galaxies are candidates for interferometric high resolution
followup observations that allow detailed studies of the spatial
distribution of molecular gas.
\
In the case of \object{UGC~1347}, such observations have been carried out using BIMA. 
\
They represent an example of how detailed investigations should be carried out on a much larger sample.
\
For this reason a description of the results is presented in Sect. \ref{UGC1347-Section}.
\\

A possible impact of the galaxies' environment on the gas distribution and star-formation properties may be investigated most easily for the brightest candidates. 
\
In order to maximize the chances of detection, the well-established CO-IR
correlation (\citealt{1986apj...310..660s}; \citealt*{1991ApJ...370..158S}) was used in the selection.   
\
All galaxies selected for observation show IRAS 100 $\mu$m fluxes of more
than 2.5 Jy, with the exception of \object{UGC~1344}, which was included
because it was discussed by \citet{2000a&a...363...41h}, together with
\object{UGC~1347}.
\
In addition to the FIR criterion, we only chose  galaxies of known HI content, as determined by \citet{1985apj...292..404g}. 
\
 
The sample was composed to contain both a subset of 7 galaxies
located close to the cluster center and a subset of 5 galaxies with a
distance to the center that exceeds the Abell radius. 
\
The results of the measurements are shown in Fig. \ref{Center-Spectra} and listed in Table \ref{CO-results} and a discussion follows in Sect. \ref{Discussion}.
\ 
\subsubsection{Molecular gas content} \label{SecMolecularMass}
\begin{figure}
  \resizebox{\hsize}{!}{\includegraphics{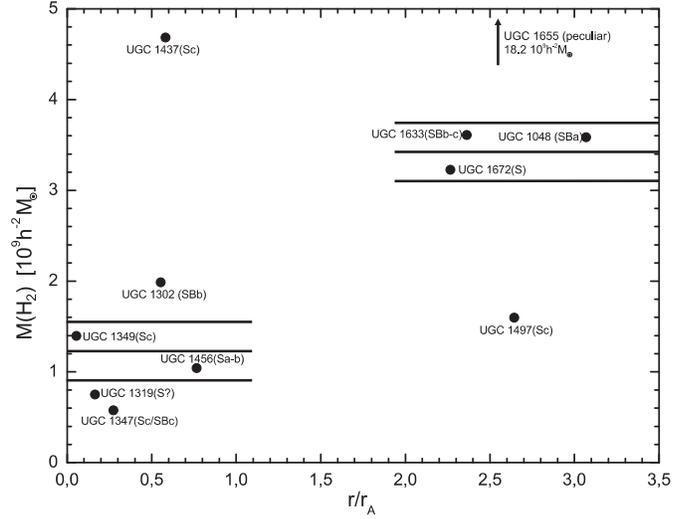}}
  \caption{Molecular mass vs. the distance to the cluster center.
  The horizontal lines represent the median value for each of the two subsamples and the common median absolute deviation. 
  To be able to compare the mass estimates of a homogeneous sample, the peculiar galaxy UGC~1655 and the FIR weak galaxy UGC~1344 were not considered (see text).  
   }     
  \label{MH2-Distance}
\end{figure}
\
The CO emission region can extend somewhat beyond the region covered by the 21.5\arcsec~IRAM 30m telescope beam at 115 GHz. This can be seen in the case of UGC~1347 in Sect. \ref{UGC1347-Section}, and it is also indicated by the off-center measurements in some other cases.
\
Therefore, to estimate the overall intrinsic CO content and to derive molecular gas masses, certain assumptions on source sizes and distributions were necessary -- namely both beam and sources were assumed to be Gaussian.
\
Please note that the need for source-size estimates introduces a serious uncertainty in the gas mass estimates. 
\
Better statistics and high resolution information indicating the actual sizes of the CO emission regions, as obtained with BIMA for the case of UGC~1347, are mandatory for verifying the results of this study.  

\
\citet{1995ApJS...98..219Y} found the CO emission region of galaxies to be confined to the center of the disk and correlated with the 25.0 B$_{\rm{mag}}$/arcsec$^2$
isophotal diameter D$_{25}$. 
\
The mean ratio of the CO scale length of a simple exponential model to D$_{25}$
was determined to be 0.10.     
\
Similar to the approach of \citet{1995ApJS...98..219Y}, the FWHM of the
sources were taken as 0.2 $\cdot$ D$_{25}$, except for the case of UGC~1347, in
which the interferometric measurements allowed a more accurate determination of
the source size. 
\
The size estimates were then used to determine beam filling factors
$F_{\rm{gauss}}=1+\left(\theta_{\rm{Beam}}/\theta_{\rm{Source}}\right)^2$
that were applied to correct the measured values of I$_{\rm{CO}}$.

\citet{1998aj....115..405l} observed UGC~1302, UGC~1319, and UGC~1456 in $^{12}$CO(1-0) line emission using the 12m NRAO telescope with a beam size of 55\arcsec and obtained values for I$_{\rm{CO}}$ of 1.4$\pm$0.21 K km s$^{-1}$ (9.68$\pm$0.76 K km s$^{-1}$ in this paper), 2.48$\pm$0.21 K km s$^{-1}$ (8.43$\pm$0.35 K km s$^{-1}$), and 2.56$\pm$0.29 K km s$^{-1}$ (9.77$\pm$0.48 K km s$^{-1}$) respectively. 
\
The differences of the values presented in this paper result from different beam sizes. 
\
Considering the different sizes by correcting with $F_{\rm{gauss}}$ yields intrinsic I$_{\rm{CO}}$ values of 6 K km s$^{-1}$ (15 K km s$^{-1}$ in this paper) for UGC~1302, 68 K km s$^{-1}$ (42 K km s$^{-1}$) for UGC~1319, and 37 K km s$^{-1}$ (31 K km s$^{-1}$) for UGC~1456. 
\
The remaining differences are explicable with uncertainties in the source-size estimates.
\
Especially UGC~1302 is likely to be resolved in the IRAS 30m telescope beam.

Determination of the H$_2$ content using I$_{\rm{CO}}$ (especially
optically thick line emission) as the tracer has to be handled with care. 
\
The common approach of using a 'standard' N(H$_2$)/I$_{\rm{CO}}$ conversion
factor X derived from galactic observations has the shortcoming of disregarding
the dependency of X on the metallicity of the region of interest
\citep{1997A&A...328..471I}. 
\
Nevertheless, since only global mass estimates can be derived and since they are caused by the lack of information on metallicity,  a 'standard' conversion factor 2.3~$\cdot
10^{20}$cm$^{-2}$ (K~km~s$^{-1}$)$^{-1}$ \citep{1988A&A...207....1S} was
adopted.
\
This also allows the results to be compared to published data.
\
The resulting molecular mass estimates are listed in Table
\ref{molecular-masses}, where a linewidth of 500 km s$^{-1}$ was assumed for the upper mass limit of UGC~1344, and plotted against their projected
distance to the cluster center in Fig. \ref{MH2-Distance}.
\
\begin{table}
  \caption[]{Molecular gas mass estimates of the sample. The sources are assumed do be Gaussian with a FWHM $\theta_{\rm{S}}$ = 0.2$\cdot$D$_{25}$, with the exception of UGC~1347, for which a more accurate FWHM could be determined from the interferometric measurements. The CO deficiency indicator CODEF follows the definition given in \citet{1998a&a...331..451c}.}. 
  \begin{tabular*}{\linewidth}{@{\extracolsep\fill} l....}
   \hline
   \hline
   \multicolumn{1}{c}{Obj.} & \multicolumn{1}{c}{$\theta_{\rm{S}}$} & \multicolumn{1}{c}{M(H$_2$)}                    & \multicolumn{1}{c}{L$_{\rm{FIR}}$/M(H$_2$)} &  \multicolumn{1}{c}{CODEF}  \\ 
                            & \multicolumn{1}{c}{[\arcsec]}         & \multicolumn{1}{c}{[10$^9 h^{-2}$  M$_{\sun}$]} & \multicolumn{1}{c}{[L$_{\sun}$/M$_{\sun}$]}  \\
      \hline
  \object{UGC 1048} & 19.0 &  3.59 & 4.4 & -0.08\\
  \object{UGC 1302} & 29.5 &  1.99 & 2.9 & 0.15 \\
  \object{UGC 1319} & 10.7 &  0.75 & 6.5 &      \\
  \object{UGC 1344} & 19.5 &  <0.65 &    &      \\
  \object{UGC 1347} & 18.9 &  0.57 & 12.6& 0.39 \\
  \object{UGC 1349} & 15.5 &  1.39 & 4.7 & -0.02\\
  \object{UGC 1437} & 30.1 &  4.68 & 4.1 & -0.06\\
  \object{UGC 1456} & 15.1 &  1.04 & 3.7 & 0.01 \\
  \object{UGC 1497} & 19.0 &  1.60 & 6.6 & 0.12 \\
  \object{UGC 1633} & 35.4 &  3.61 & 2.9 & 0.06 \\
  \object{UGC 1655} & 34.6 &  18.2 & 2.7\\
  \object{UGC 1672} & 13.2 &  3.23 & 8.6\\
  \hline
 \end{tabular*}
 \label{molecular-masses}     
\end{table}
\
In this plot the horizontal lines represent the median molecular gas estimates
and the common median absolute deviation for the two subsamples.
\
To compare a more homogenous set of galaxies, the peculiar galaxy UGC~1655 with its outstanding flux density was not considered in the calculation of the median. 
\
Nor was UGC~1344, which does not meet the 100$\mu$m selection criterion.
\
If both galaxies were included, the median values of the two subsamples would be even further separated.
\

\subsubsection{L$_{\rm{FIR}}$ properties}
\label{LFIR-Results}
\
\begin{table*}
  \caption[]{Properties of the IRAS FIR flux limited sample of Abell 262  galaxies with a distance to the cluster core of less than 3 Abell radii, as observed in HI by \citeauthor{1985apj...292..404g} .  The morphological classification, as well as the inclination estimate (with respect to the line of sight), were taken from 
  \citet{1973UGC...C...0000N}, D$_{25}$ from \citet{1991trcb.book.....D},
  S$_{60\mu \rm{m}}$ and S$_{100\mu \rm{m}}$ IRAS flux densities from
  \citet{1990irasf.c......0m}, and HI deficiency from
  \citet{1985apj...292..404g}. An inclination value of 1 indicates a face-on, and a 7 indicates an edge-on view onto the galaxy.}     
  \begin{tabular*}{\linewidth}{@{\extracolsep\fill}lll........}
     \hline
     \hline
  \multicolumn{1}{c}{Obj.} &   \multicolumn{1}{c}{Class} & \multicolumn{1}{c}{Incl.} &\multicolumn{1}{c}{D$_{25}$}  &\multicolumn{1}{c}{S$_{12\mu \rm{m}}$}  &\multicolumn{1}{c}{S$_{25\mu \rm{m}}$} &\multicolumn{1}{c}{S$_{60\mu \rm{m}}$} &\multicolumn{1}{c}{S$_{100\mu \rm{m}}$} &\multicolumn{1}{c}{L$_{\rm{FIR}}$}             & \multicolumn{1}{c}{HI}        & \multicolumn{1}{c}{r/r$_A$} \\ 
                           &                             &  class                         &\multicolumn{1}{c}{[\arcmin]} &\multicolumn{1}{c}{[Jy]}                &\multicolumn{1}{c}{[Jy]}               &\multicolumn{1}{c}{[Jy]}               &\multicolumn{1}{c}{[Jy]}                &\multicolumn{1}{c}{[10$^9 h^{-2}$ L$_{\sun}$]} & \multicolumn{1}{c}{deficiency}&                             \\
      \hline
               
  UGC 1048 &  SBa      & 1  &  1.58 & 0.14& 0.47  & 3.75 &  6.95  &   15.76 &   0.2 & 3.07\\
  UGC 1089 &  Sc       & 1  &  1.15 & 0.12& 0.19  & 1.08 &  3.17  &    5.65 &  0.14 & 2.88\\
  UGC 1094 &  SBb      & 6  &  2.24 & 0.11& 0.16  & 1.03 &  3.19  &    5.55 &  0.07 & 2.9\\
  UGC 1100 &  Sb/SBb   & 6  &  2.19 &<0.11&<0.08  & 0.25 &  1.02  &    1.57 &  0.29 & 2.41\\
  UGC 1125 &  S        &    &       &<0.08&<0.12  & 0.42 &  1.05  &    2.02 &  -0.2 & 2.44\\
  UGC 1152 &  Sb       & 5  &  1.05 &<0.08&<0.10  & 0.32 &  0.79  &    1.53 & -0.12 & 3.05\\
  UGC 1178 &  Sc       & 7  &  1.82 & 0.14& 0.12  & 1.23 &  3.66  &    6.48 & -0.27 & 1.69\\
  UGC 1220 &  S        &    &  0.83 & 0.14& 0.26  & 1.88 &   4.6  &    8.95 &     0 & 1.3\\
  UGC 1234 &  Sc/SBc   &    &  1.12 &<0.07&<0.09  & 0.23 &  0.56  &     1.1 & -0.14 & 1.01\\
  UGC 1238 &  Sb       & 3  &  1.78 & 0.08& 0.12  & 0.75 &  2.05  &    3.78 &  0.03 & 0.77\\
  UGC 1248 &  Sa-b     & 6  &  3.16 & 0.09&<0.12  & 0.28 &  1.63  &    2.24 &  0.77 & 0.73\\
  UGC 1251 &  peculiar &    &  0.91 &<0.10& 0.18  & 0.03 &  0.55  &    0.59 &  0.83 & 0.62\\
  UGC 1302 &  SBb      & 3  &  2.45 & 0.11& 1.29  & 1.29 &  2.78  &    5.79 &   0.1 & 0.55\\
  UGC 1319 &  S?       &    &  0.89 & 0.12&<0.20  & 0.92 &  2.74  &    4.85 &  0.27 & 0.16\\
  UGC 1338 &  Sb       & 2  &  0.89 &<0.13&<0.13  & 0.29 &  0.85  &    1.52 &>0.79  & 0.21\\
  UGC 1344 &  SBa      & 5  &  1.62 &     &       &      &        &         &>0.78  & 0.21\\
  UGC 1347 &  Sc/SBc   & 1  &  1.26 & 0.12& 0.19  & 1.49 &  3.84  &    7.28 &-0.07  & 0.27\\
  UGC 1349 &  Sc       & 1  &  1.29 & 0.14& 0.21  & 1.44 &  3.26  &    6.61 & 0.38  & 0.05\\
  UGC 1358 &  S0-a     & 6  &  1.55 & 0.07&<0.04  & 0.09 &  0.31  &    0.51 &>0.79  & 0.09\\
  UGC 1376 &  SBb      & 4  &   1.7 &<0.10&<0.11  & 0.26 &   0.9  &    1.48 & 0.22  & 1.88\\
  UGC 1411 &  Sb       & 6  &  1.82 &<0.11&0.01   & 0.44 &  1.45  &    2.46 & -0.3  & 1.22\\
  UGC 1421 &  S        & 7  &  1.62 &<0.08&<0.09  &  0.3 &  0.98  &    1.67 &  0.48 & 1.86\\
  UGC 1422 &           &    &  1.17 &<0.09&<0.11  & 0.31 &  0.87  &     1.6 &  0.17 & 2.00\\
  UGC 1437 &  Sc       & 3  &  2.51 & 0.29& 0.38  & 3.36 &  11.4  &   19.04 & -0.15 & 0.58\\
  UGC 1456 &  Sa-b     & 3  &  1.26 & 0.09&<0.18  & 0.62 &   2.5  &     3.9 &  0.46 & 0.77\\
  UGC 1493 &  SB?a-b   & 6  &  1.82 & 0.16& 0.16  & 1.29 &  3.65  &    6.61 &  0.75 & 1.51\\
  UGC 1497 &  Sc       & 2  &  1.58 & 0.17& 0.29  & 2.02 &  5.85  &   10.49 & -0.09 & 2.64\\
  UGC 1520 &  peculiar &    &  0.63 & 0.14&<0.22  & 1.67 &  4.07  &    7.95 & -0.08 & 2.61\\
  UGC 1550 &  Sc       & 6  &  3.16 & 0.13& 0.22  & 1.45 &  5.06  &    8.34 &   0.1 & 1.74\\
  UGC 1581 &  S-Irr    &    &  1.66 &<0.11& 0.80  &  0.8 &   2.4  &    4.22 & -0.29 & 1.66\\
  UGC 1633 &  SBb-c    & 6  &  2.95 & 0.21& 0.25  & 1.91 &   6.2  &   10.55 &  0.21 & 2.36\\
  UGC 1655 &  peculiar &    &  2.88 & 0.70& 1.03  &10.89 & 22.83  &    48.3 &  0.25 & 2.64\\
  UGC 1672 &  S        & 6  &   1.1 & 0.37& 0.75  & 6.43 & 12.54  &   27.62 & -0.06 & 2.27\\
  UGC 1676 &  Sba/Sb   & 5  &  1.78 &<0.10&<0.12  & 0.86 &   2.6  &    4.57 &  0.06 & 2.26\\
  UGC 1721 &  SBb/SBc  & 1  &     2 &<0.11&<0.12  & 0.19 &  0.77  &     1.2 & -0.14 & 2.61\\
  UGC 1769 &  Sb/Sc    & 4  &     1 &0.08 &<0.14  & 0.68 &  2.33  &    3.86 & -0.31 & 2.98\\
  Z522-002 &           &    &  0.74 &<0.10&<0.07  &  0.3 &  0.71  &     1.4 & -0.21 & 1.02\\
  \hline
\end{tabular*}                            
\label{Abell262-Properties}
\end{table*}
\
\begin{figure}[h]
  \resizebox{\hsize}{!}{\includegraphics{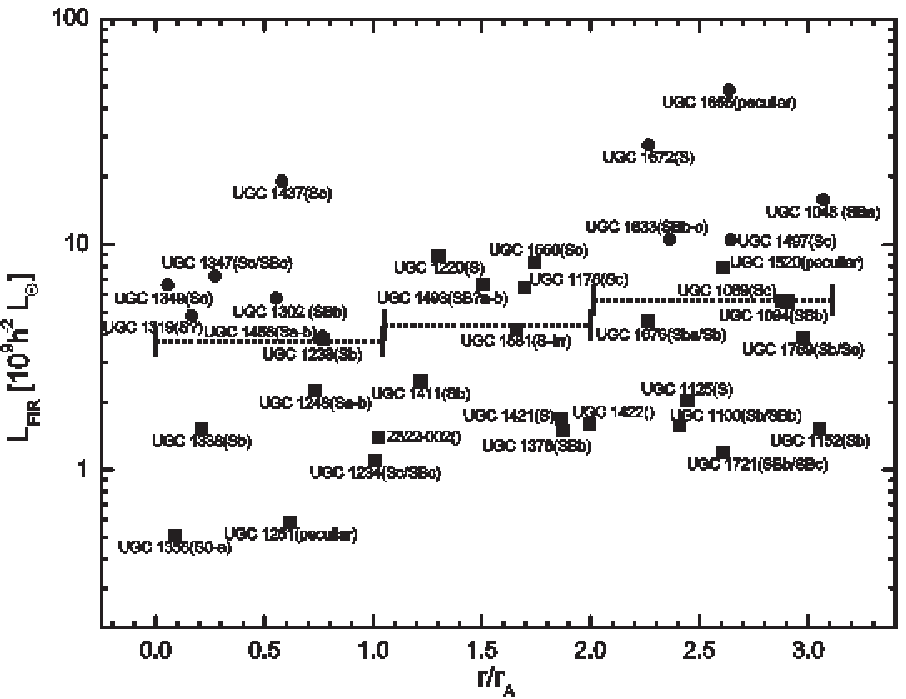}}
  \caption{FIR luminosity as function of the distance to the cluster center. The plot contains all Abell~262 galaxies down to the IRAS flux limit, with distances to the cluster core not exceeding 3 Abell radii. The filled circles represent the objects for which CO data were obtained. Median values of 3 regions are shown by the horizontal lines.}     
  \label{LFIR-Distance}
\end{figure}
\
To be able to discuss a possible selection bias due to the FIR selection criterion, FIR luminosities were calculated for all  Abell~262 galaxies in the \citeauthor{1985apj...292..404g} sample within 3 Abell radii down to the IRAS flux limit (Table \ref{Abell262-Properties}).
\
Following \citet{1988apjs...68..151h} the FIR luminosity can be written as:
$$\rm{L}_{\rm{FIR}} = 3.94\cdot 10^5 \rm{D[Mpc]}^2 (2.58 \cdot\rm{S}_{60\mu \rm{m}}\rm{[Jy]}+\rm{S}_{100\mu \rm{m}}\rm{[Jy]}).$$ 
In Fig. \ref{LFIR-Distance} it is plotted against the projected distance to the cluster center.  
\
\subsection{Interferometric observations of UGC 1347}
\label{UGC1347-Section}
\begin{figure}
  \resizebox{\hsize}{!}{\includegraphics[angle=-90]{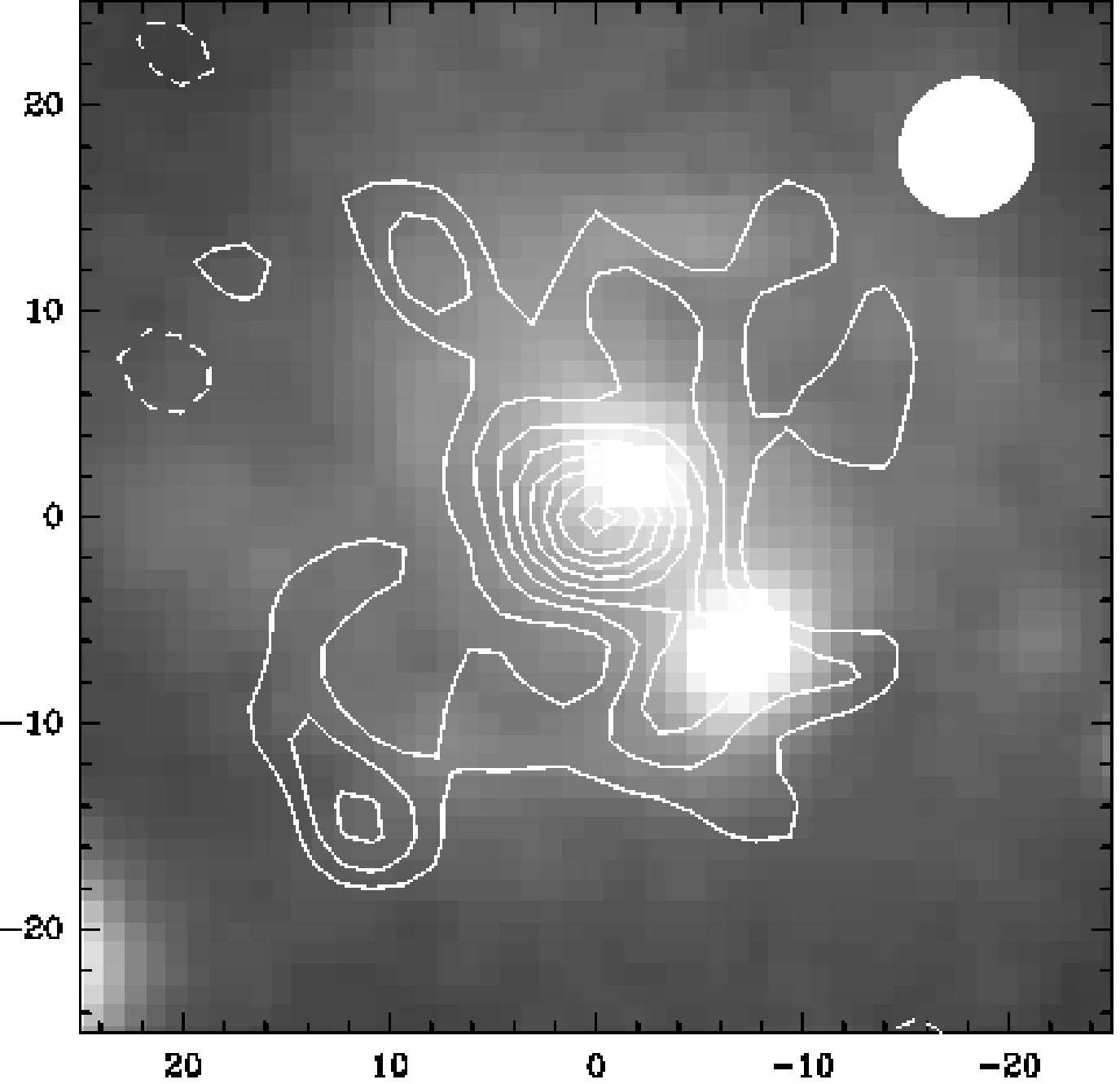}}
  \caption{Overlay of the \element[][12]{CO} map obtained with BIMA
   and the POSS-II Red image. The contours have an increment of 1$\sigma$= 1.80 Jy beam$^{-1}\cdot$ km s$^{-1}$  starting at  2$\sigma$. The CO emission extends over most of the inner disk of UGC~1347.}  
  \label{DSS-BIMA_map}
\end{figure}

\begin{figure}
  \resizebox{\hsize}{!}{\includegraphics[]{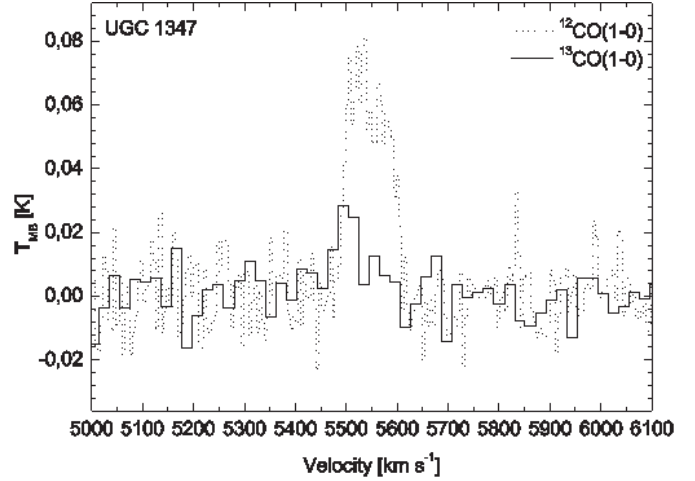}}
  \caption{Global \element[][12]{CO}(1-0) and \element[][13]{CO}(1-0) spectra
  of UGC~1347, obtained with the IRAM 30m telescope}             
  \label{UGC1347_CO10-spectra}
\end{figure}

\begin{figure}
 \resizebox{\hsize}{!}{\includegraphics[angle=-90]{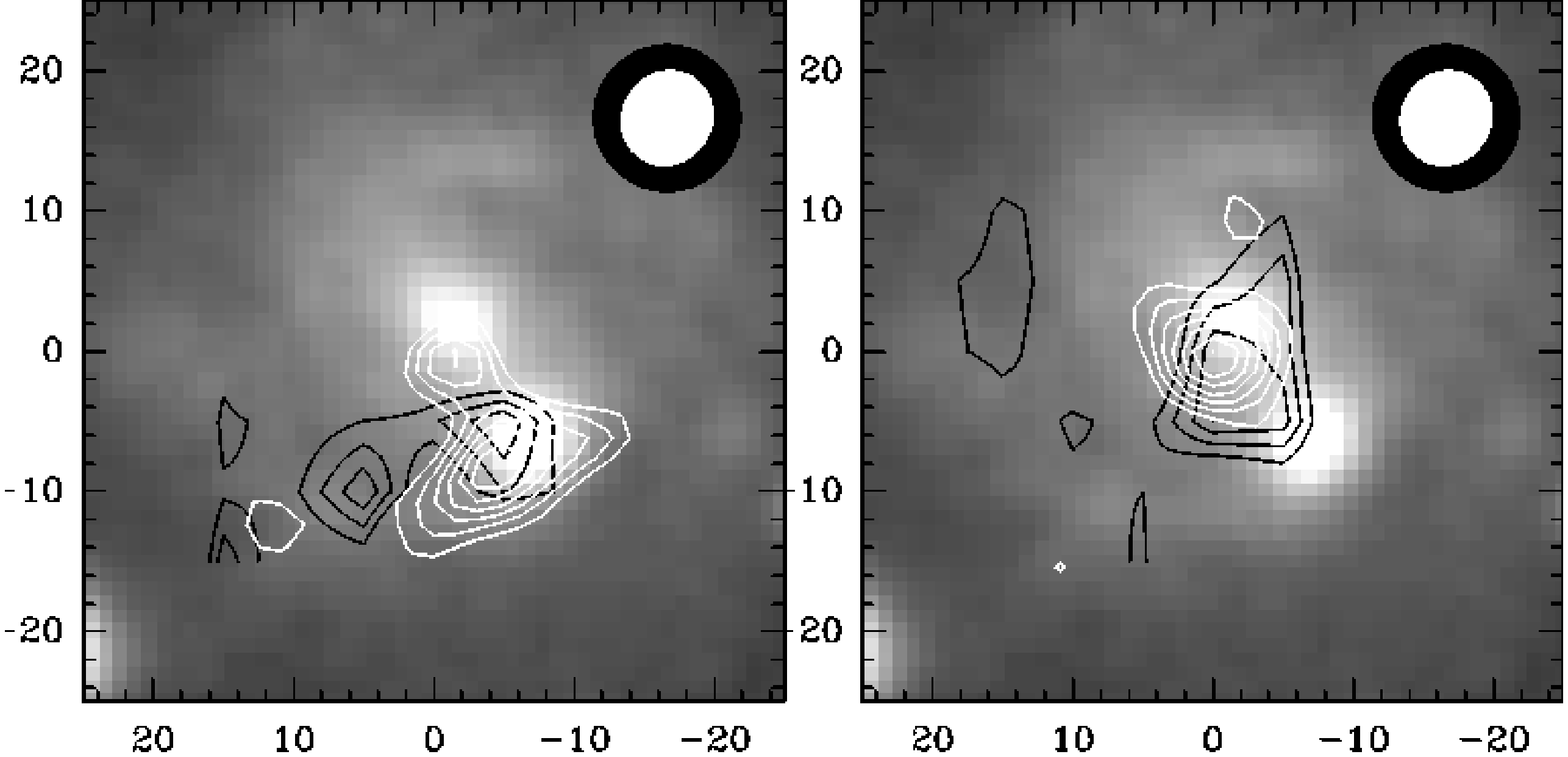}}
  \caption{Overlay of the \element[][12]{CO}(1-0) maps obtained with BIMA
  (white contours), the \element[][12]{CO}(2-1) maps obtained with the IRAM 30m
  telescope (black contours), and the POSS-II Red image. The integrated intensity maps cover the velocity ranges in which at least a 2$\sigma$ flux density contribution to the southern (left panel) and the central (right panel) source was detected with BIMA. Contours represent 10\% steps of the peak emission.}    
  \label{DSS-BIMA-IRAM_sources-maps}
\end{figure}

The overlay of the POSS-II Red\footnote{
The Second Palomar Observatory Sky Survey (POSS-II) was made by the California Institute of Technology with funds from the National Science Foundation, the National Geographic Society, the Sloan Foundation, the Samuel Oschin Foundation, and the Eastman Kodak Corporation.
} image and the BIMA CO map in Fig. \ref{DSS-BIMA_map} clearly shows that CO emission extends over almost the entire inner disk of UGC 1347. 
\
Continuum emission was not detected.
\
A comparison of the two CO(1-0) spectra can be found in Fig. \ref{UGC1347_CO10-spectra}.

With the high resolution \element[][12]{CO}(1-0) observations obtained with BIMA, it was possible to identify reservoirs of molecular gas at the positions of the nucleus and in the southern component.
\
Figure \ref{DSS-BIMA-IRAM_sources-maps} shows maps of the two components. 
\
Only channel maps with a contribution of at least 2$\sigma$ to the individual components were used for compiling the integrated intensity maps. 
\
This resulted in corresponding velocity ranges of 5464-5589 km s$^{-1}$ for the
central component, 5555-5613 km s$^{-1}$ for the southern component, and
5456-5613 km s$^{-1}$ for the total galaxy (Fig. \ref{DSS-BIMA_map}).
\
The IRAM 30m \element[][12]{CO}(2-1) flux densities were integrated over the same velocity ranges as the BIMA maps.
\
To coincide better  with the BIMA contour map and the POSS-II Red data, the IRAM maps were shifted by 1.6\arcsec~ towards north-east -- a correction which  agrees with the pointing accuracy of the 30m telescope.

Table \ref{CO-content} lists the molecular line emission data determined for
both regions, based on the BIMA \element[][12]{CO}(1-0) and the IRAM 30m \element[][12]{CO}(2-1) observations.
\
Their sizes were determined in the BIMA integrated intensity maps (Fig.
\ref{DSS-BIMA-IRAM_sources-maps}, white contours)  assuming a
Gaussian brightness distribution with an additional underlying disk, both contributing to the total flux density.   
\
For the central region, the 65\% peak intensity contour was chosen, while for the southern component the 60\% contour was chosen to represent the FWHM of the distribution.
\
The nucleus remains unresolved at the BIMA beam with a  FWHM $\sim$ 6.6\arcsec,~ whereas the southern region shows extended CO emission at this resolution. 
\
To consider the error caused by the uncertainty in the determination of the
size of the southern component, the 50\% and 70\% contours were taken as upper
and lower limits. 

The derived intrinsic \element[][12]{CO}(2-1)/\element[][12]{CO}(1-0) line
ratios R are listed in Table \ref{Lineratios}, together with the observed
line ratios R$_{\rm{Obs}}$ and the corresponding correction factors
$f_{\rm{gauss}}$, where
$$\rm{R}=f_{\rm{gauss}}\cdot \rm{R}_{\rm{Obs}};\;\;
f_{\rm{gauss}}=\frac{1+(10.5"/\Theta_{\rm{S}})^2}{1+(6.6"/\Theta_{\rm{S}})^2}$$      
and $\Theta_{\rm{S}}$\ represents the FWHM of the source.
In both regions, the CO lines are optically thick -- under normal conditions an
indication of cold, subthermally excited and dense ($\la 10^4$~cm$^{-3}$, \citealt{1990ApJ...348..434E})  molecular gas.

Due to the lack of any information on metallicity in UGC~1347, the estimate of
the molecular gas mass M(${\rm H_2}$), as listed in Table \ref{GasMass}, has a significant amount of uncertainty.  
\
To account for the higher metallicity in galaxy centers, the value of
M(${\rm H_2}$) for the central region given in Table \ref{GasMass} 
was based on a lower conversion factor than the other two values, as suggested
in \citet{2001mhs..conf..293I}. 
\

\begin{table}
  \caption[]{Integrated temperatures for both CO-bright components and for the
  disk of UGC 1347. Both components were assumed to be Gaussian. A disk flux
   contribution was taken into account in the determination of the source
   sizes. The central component is not resolved. The values for the total
   galaxy represent IRAM 30m single-beam measurements centered on the galaxy. 
   A comparison of the two CO(1-0) spectra can be found in Fig. \ref{UGC1347_CO10-spectra}.
   }      
  \begin{tabular*}{\linewidth}{@{\extracolsep\fill}lllll}
  \hline
  \hline
     & \multicolumn{1}{c}{FWHM}                   & \multicolumn{1}{c}{\element[][12]{CO}(1-0)} & \multicolumn{1}{c}{\element[][12]{CO}(2-1)} & \multicolumn{1}{c}{\element[][13]{CO}(1-0)}  \\
     & \multicolumn{1}{c}{$\Theta_{\rm{S,Obs}}$}  & \multicolumn{1}{c}{I$_{\rm{CO}}$ }          & \multicolumn{1}{c}{I$_{\rm{CO}}$}           & \multicolumn{1}{c}{I$_{\rm{CO}}$}        \\ 
     & \multicolumn{1}{c}{[\arcsec]}              & \multicolumn{1}{c}{[K km s$^{-1}$]}         & \multicolumn{1}{c}{[K km s$^{-1}$]}         & \multicolumn{1}{c}{[K km s$^{-1}$]}          \\
  \hline
  nucleus      &    $<$6.6                   & 40.1 $\pm$ 1.3        & 8.4 $\pm$ 1.0\\
  south.       &       8.5$^{+1.4}_{-1.6}$   & 24.1$^{+5.5}_{-0.6}$  & 4.0 $\pm$ 0.6 \\
  total galaxy &     18.9                      & 4.6 $\pm$ 0.3         &   &  1.7$\pm$0.3\\
  \hline
  \label{CO-content}
  \end{tabular*}
\end{table}

\begin{table}
  \caption[]{\element[][12]{CO}(2-1)/\element[][12]{CO}(1-0) line ratio R for
  both CO-bright components of UGC 1347. The listed ratios R and source sizes
  $\theta_{\rm{S}}$ are intrinsic values based on the integrated temperatures
  and observed sizes given in Table \ref{CO-content} and assuming Gaussian
  distributions for the beams and  sources.}       
  \begin{tabular*}{\linewidth}{@{\extracolsep\fill}lllll}
  \hline
  \hline
     &\multicolumn{1}{c}{FWHM $\theta_{\rm{S}}$ [\arcsec]} & \multicolumn{1}{c}{R$_{\rm{Obs}}$}  & \multicolumn{1}{c}{$f_{\rm{gauss}}$}  &  \multicolumn{1}{c}{R}\\
  \hline
  nucleus & 6.6 & 0.21 $\pm$ 0.03 & 1.77 & 0.37\\
          & 0   &                 & 2.50 & 0.53\\
  south.  & 5.4$^{+2.0}_{-3.4}$ & 0.17$^{-0.03}_{+0.00}$  & 1.92$^{-0.24}_{+0.48}$ & 0.33$^{-0.09}_{+0.08}$\\
    \hline
  \label{Lineratios}
  \end{tabular*}
\end{table}
\
\begin{figure}
  \resizebox{\hsize}{!}{\includegraphics[]{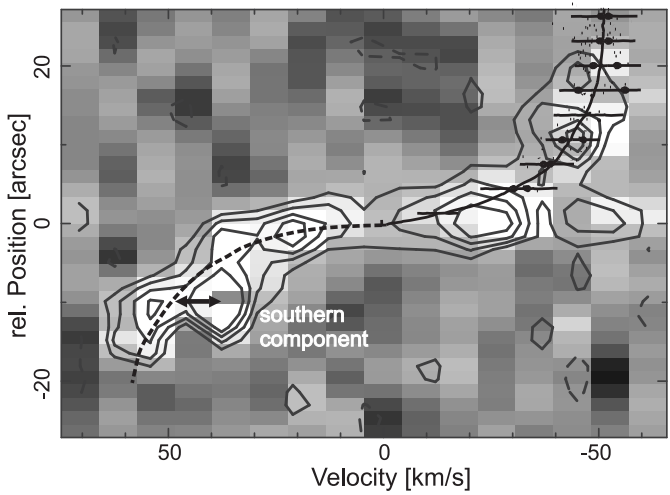}}
  \caption{\element[][12]{CO}(1-0) position-velocity diagram of UGC 1347 with
  the origin in the center of the galaxy and a position angle of 20\degr.
  Superimposed in the top right corner is the H$\alpha$ rotation curve obtained
  by \citet{1994a&as..103....5a} (see text).  
  The southern component is blue-shifted by approximately 10 km s$^{-1}$ with respect to the disk radiation at that position (indicated by the arrow).}          
  \label{RotationCurve}
\end{figure}
\
\begin{table}
  \caption[]{Molecular gas masses of the nucleus, southern component, and total
  galaxy. The size estimates given in Table \ref{Lineratios} were applied.}           
  \begin{tabular*}{\linewidth}{@{\extracolsep\fill}lll}
  \hline
  \hline
     &\multicolumn{1}{c}{X } & \multicolumn{1}{c}{M(${\rm H_2}$)} \\
     &\multicolumn{1}{c}{[cm$^{-2}$ (K km s$^{-1}$)$^{-1}$]} & \multicolumn{1}{c}{[M$_{\sun}$]} \\
  \hline
  nucleus      & 4$\cdot 10^{19}$ & 4.7$\cdot 10^7 h^{-2}$\\
  south.       & 2$\cdot 10^{20}$ & 9.4$\cdot 10^7 h^{-2}$\\
  total galaxy & 2$\cdot 10^{20}$ & 5.7$\cdot 10^8 h^{-2}$\\
  \hline
  \label{GasMass}
  \end{tabular*}
\end{table}

\citet{1994a&as..103....5a}  investigate the H$\alpha$ velocity field and the
H$\alpha$ rotation curve of UGC~1347. They derive an approximate inclination of 30\degr,~ as well as a  position angle of 20\degr.
\
As shown in Fig. \ref{RotationCurve}, the H$\alpha$ rotation curve is in good
agreement  with the \element[][12]{CO}(1-0) velocity pattern along the same
axis.  
\
The H$\alpha$  rotation curve was scaled by a factor of 0.5 along the velocity axis to compensate for the inclination correction that was applied to the \citet{1994a&as..103....5a} data.
\
However, the coincidence of both curves is valid only relative to each other, since \citeauthor{1994a&as..103....5a} find a systemic velocity of only 5478 km s$^{-1}$.

Another remarkable feature visible in the position-velocity diagram is a
discrepancy in the velocity of the southern component and the disk at the
same position.
\
The gas appears to be blue-shifted by approximately
10 km s$^{-1}$ with respect to the underlying disk.
\
A weak evidence of a component, which is to a lesser extent red-shifted at that position is also indicated in the diagram.
\
This may be explained by an expected cloud-cloud velocity dispersion of $\sigma\la$10 km s$^{-1}$.
\
However, the blue-shifted (as well as the indicated red-shifted) emission coincides with the position of the southern component. Therefore, the shift may alternatively be explained by the model of an asymmetrically expanding gas shell or of large scale outflow with a substantial amount of entrained molecular gas. 
\
It supports the hypothesis of enhanced star formation in that region, as indicated by \cite{2000a&a...363...41h}.

As expected, the largest velocity dispersion can be found at the position of the nuclear source.
\
An upper limit for the dynamical mass confined in the unresolved center can be derived by considering the maximum velocity measured at the center position in the position-velocity diagram of $\sim$ 50km s$^{-1}$. 
\
When accounting for an inclination of 30\degr~ (resulting in v$_{\rm{max}}$=100km
s$^{-1}$) and assuming a maximum radius of below 1.5\arcsec, an upper limit of 8.3$\cdot$10$^8$$h^{-1}$ M$_{\sun}$ can be derived.

\section{Discussion}
\label{Discussion}
In the following the question about possible environmental influence on the global molecular mass content of cluster galaxies is revisited. In Sect. \ref{SecH2LossSignatures} we present several indicators of the presence of a molecular gas deficiency in the cluster core. This deficiency is discussed further in the framework of a cirrus model in Sect. \ref{SecCirrusStripping}. Section \ref{SecCODEF} comments on a previously introduced CO deficiency parameter and in Sect. \ref{Morphology} the morphological composition of the subsamples of Abell~262 is discussed.        
\
\subsection{Signatures of environmental influences on the global molecular mass content}
\label{SecH2LossSignatures}
The molecular gas mass estimates shown in Fig. \ref{molecular-masses} indicate a
dependency of the total molecular gas content in Abell~262 galaxies on the projected distance to the cluster core.  
\
The median molecular gas mass of the sample members located in the vicinity of the cluster center was found to be $1.22 \pm 0.55 \cdot 10^9 h^{-2} \rm{M}_{\sun}$, whereas a value of $3.41 \pm 0.20 \cdot 10^9 h^{-2} \rm{M}_{\sun}$ was  determined  for the more distant subsample.
\
The difference in the two median values is $\sim4$ times larger than the
median absolute deviations.

So far, data exist only for the brightest FIR galaxies in the cluster. 
With the caveat of a small statistical base, such a trend must be restricted to these objects. 
\
For a more general statement, a larger amount of galaxies has to be targeted in upcoming CO observations. 
\
Figure \ref{LFIR-Distance} shows that the application of the FIR selection criterion results in the inclusion of galaxies (such as the peculiar galaxy UGC~1655, UGC~1672, or UGC~1437) with outstanding L$_{\rm{FIR}}$ compared to the bulk of IRAS galaxies in Abell~262. 
\
The strong FIR flux density of UGC~1437, the location within the IRAS color-color diagram (cf. Sect. \ref{SecCirrusStripping}), and the fact that it is not HI deficient, together indicate that this galaxy is a more distant object that is projected into the central region.

Other indicators can be found in the L$_{\rm{FIR}}$ data that support the presence of an environmental influence on the molecular gas content of cluster galaxies. 
\
With the exception of UGC~1437 mentioned above, the L$_{\rm{FIR}}$ brightest Abell~262 galaxies are found at distances that exceed one Abell radius, whereas the faintest sources detected by IRAS are located close (in projection) to the core of the cluster. 
\
\begin{figure}
  \resizebox{\hsize}{!}{\includegraphics[]{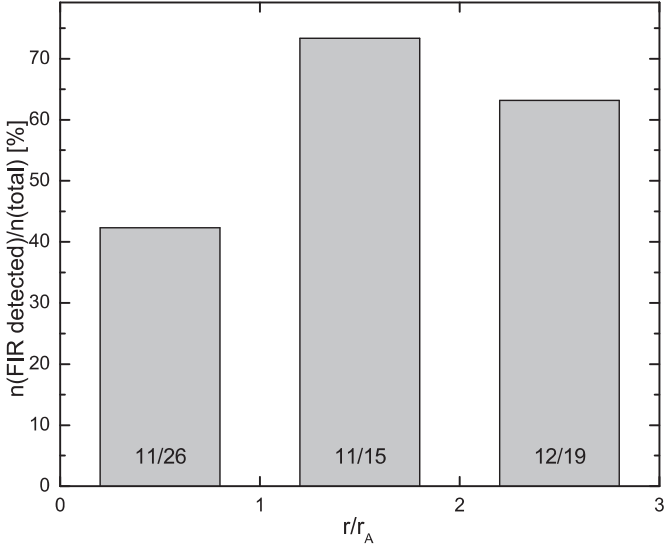}}
  \caption{The ratio of galaxies detected by IRAS  varies with the distance to the cluster core.}             
  \label{LFIR-Detection-Ratio}
\end{figure}
Moreover, the ratio of sources detected by IRAS in the Abell~262 sample of  \citeauthor{1985apj...292..404g} drops from $\sim$65\% within the region of the annulus from 1 to 3 Abell radii to $\sim$45\% in the central region of the cluster (cf. Fig. \ref{LFIR-Detection-Ratio}).
\
Due to projection effects, the intrinsic fraction of FIR-detected sources in the center ($n_{\rm{c}}$) appears to be overestimated.
\
There $n_{\rm{c}} = n_{\rm{total}}-n_{\rm{fg}}$, where $n_{\rm{total}} (=26)$ represents all galaxies found within the central Abell radius, including a number of foreground or background galaxies $n_{\rm{fg}}$  that are projected along the line of sight into the central region. 
\
This number can be approximated by the product of the surface number density of galaxies of the outer region and the area of the central region:     
\      
$$n_{\rm{fg}} \approx  \frac{n_{\rm{outer}}} {r_{\rm{outer}}^2 - r_{\rm{inner}}^2}  \cdot r_{\rm{inner}}^2  $$ 
\
where $n_{\rm{outer}}$ (=34) represents the number of galaxies within the projected region between 1 Abell radius ($r_{\rm{inner}}$) and 3 ($r_{\rm{outer}}$) Abell radii.
\
This approximation leads to an upper limit for the number $n_{\rm{c}}$ of galaxies within $r_{\rm{inner}}$, since the cluster profile has not been taken into account. 
\
The application of this correction yields a ratio of IRAS detected sources to $n_{\rm{c}}$ of below 38\% within the cluster core in comparison with $\sim$65\% in the outer regions.

Aside from the IRAS detection rate, the median L$_{\rm{FIR}}$ value of detected galaxies in Abell~262 also varies with the projected distance to the cluster core.
\  
While in the region between 2 and 3 Abell radii, the median value amounts to $5.6\cdot10^9$L$_{\sun}$, it drops to $3.8\cdot10^9$L$_{\sun}$ (cf. Fig. \ref{LFIR-Distance}) in the region of the central Abell radius. 
\
Again, the intrinsic value is likely to be lower, since more distant galaxies with higher L$_{\rm{FIR}}$ values are projected into the core region.

Assuming the validity of the M(H$_2$)--L$_{\rm{FIR}}$ correlation, the drop in the IRAS detection rate and median L$_{\rm{FIR}}$ value in the center may imply a loss of molecular gas of galaxies in the core region of the cluster.
\
This tentative result is not expected from previously published studies of molecular gas in galaxy clusters.
\
However, our results are not necessarily inconsistent with previous investigations.
\
The existing studies are rather inhomogeneous in terms of selection criteria and focus. Therefore the results are difficult to compare. 
\
A  conclusion common to all studies is  that the influence of the environment on the molecular gas content of cluster galaxies is minor (in many cases insignificant) and not comparable to the environmental impact on HI abundances. 
\
\citet{1989apj...344..171k} study the CO content of a blue-magnitude limited sample of Virgo galaxies as a function of the HI deficiency of these objects. 
\
\citet{1991a&a...249..359c} take a similar approach and analyse the CO content of a FIR-selected sample of 9 HI-deficient Coma galaxies with 9 HI normal Coma galaxies.  
\ 
\citet*{1995A&A...303..361H} focus on the fact that  FIR or blue-magnitude selected samples are biased towards strong CO emission. They detected 11 of 21 Fornax cluster galaxies in CO emission.
\
All 3 studies suffer from a very limited number of observations.

In contrast, \citet{1997a&a...327..522b} present a study on an optically selected sample of 27 galaxies in the Coma and Abell~1367 clusters and compare their CO content with a sample of 37 relatively isolated galaxies.
\
Their search for a CO deficiency of cluster galaxies as a function of the distance from the cluster center shows a higher dispersion in the central region, thus leaving room for a minor environmental influence on the CO content.
\
With a total of 582 galaxies  from their own observations and from the literature, the study of \citet{1998a&a...331..451c} is based on the broadest statistical base. 
\
Their definition of the CO deficiency is discussed in the following.

\subsection{The CODEF parameter}
\label{SecCODEF}
\begin{figure*}
  \centering
  \includegraphics[width=8cm]{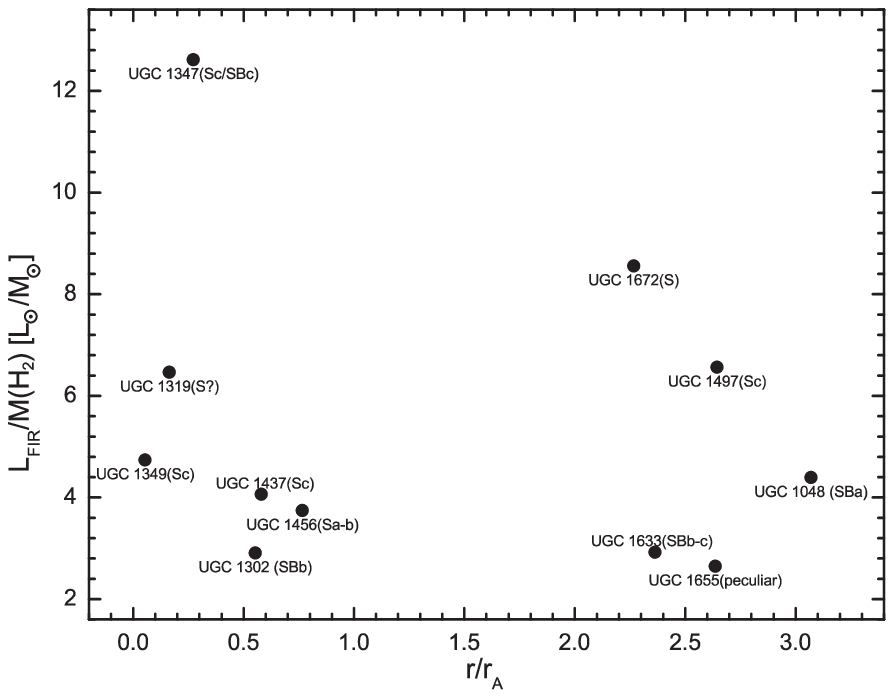}
  \includegraphics[width=8cm]{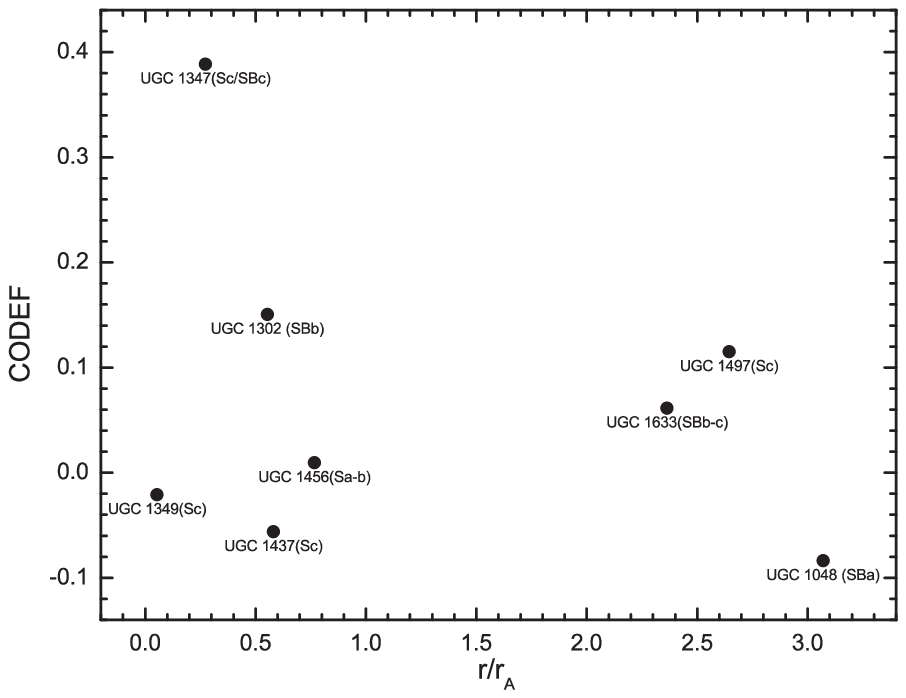}
  \caption{Star-formation efficiency and CODEF parameter as a function of the distance to the cluster core.}    
  \label{SFE_dist}\label{CODEF_dist}
\end{figure*}
\
Similar to the definition of the HI deficiency in \citet{1985apj...292..404g},  
\citet{1998a&a...331..451c} propose a CO deficiency parameter: 
$$ \rm{CODEF} = \log((\rm{M}({\rm{H}_2})/\rm{D}_{25}^2)_{\rm{expected}}) - \log((\rm{M}({\rm{H}_2})/\rm{D}_{25}^2)_{\rm{obs}}).$$
\
As before, $\rm{M}({\rm{H}_2)}$ is assumed to be directly proportional to I$_{\rm CO}$. 
\
In order to compare the CO emission of observed galaxies with values obtained for isolated reference galaxies, \citet{1998a&a...331..451c} introduce a CO-emission predictor, which makes use of the FIR to optical-size relation $\log(\rm{L}_{\rm{FIR}}/\rm{D}_{25}^2$):
\
$$\log((\rm{M}({\rm{H}_2})/\rm{D}_{25}^2)_{\rm{expected}}) = \rm{a}(T) \log(\rm{L}_{\rm{FIR}}/\rm{D}_{25}^2) + \rm{b}(T),$$ 
where a and b depend on the morphological type.
\
The morphological type and optical size are already used by \citet{1984aj.....89..758h} to compare the HI content of galaxies to a standard of normalcy for isolated galaxies.
\
In addition to these two parameters, \citet{1998a&a...331..451c} use $\rm{L}_{\rm{FIR}}$ in their extended standard of normalcy for the molecular gas content of isolated galaxies. 
\
However, the use of $\rm{L}_{\rm{FIR}}$ has the shortcoming of being correlated with the CO content of a galaxy. 
\
A CO deficiency may result in reduced FIR emission, as indicated in Sect. \ref{SecH2LossSignatures}.
\
In this case, the CO emission predictor will yield  lower values for the `expected' normalized CO emission  $\log((\rm{M}({\rm{H}_2})/\rm{D}_{25}^2)_{\rm{expected}})$ than what it ideally should, which in turn reduces the value of CODEF.   
\
In other words,  $\rm{L}_{\rm{FIR}}$ is well known for not being orthogonal to the CO emission. Therefore, this parameter is not  well-suited to defining a standard of normalcy in a CO-deficiency study.

In the way it is defined above, CODEF instead forms a morphology dependent parameter that is similar to the star-formation efficiency $\rm{L}_{\rm{FIR}}/M_{\rm{H}_2}$ (cf. Fig. \ref{CODEF_dist} and  \citet{1988apj...334..613s} and references therein). 
\
The SFE can be regarded as an indicator of the morphological type of galaxies and, therefore, is likely to be an indirect measure of their overall molecular gas content rather than their deficiency.

The CODEF values for the Abell~262 sample galaxies of known galaxy type are shown in Table \ref{molecular-masses} and plotted in Fig. \ref{CODEF_dist}. 
\
A Hubble constant H$_0$=75km s$^{-1}$ Mpc$^{-1}$ was assumed to comply with the \citet{1998a&a...331..451c} parameters. 
\
Because of the restriction to non-peculiar galaxies, the number of galaxies in our sample for which CODEF can be derived is even smaller than the number of galaxies with measured CO-line intensities.  
\
For the remaining galaxies, a dependency of CODEF on the distance to the cluster core cannot be identified.
\subsection{Morphological distribution}
\label{Morphology}
\
A more detailed analysis of a possible influence of the environment on the global molecular gas content has to include the morphological distribution of galaxies within the cluster.
\
The \citet{1985apj...292..404g} sample of Abell~262 galaxies indeed shows  segregation between Sa/b and Sc galaxies as a function of position within the cluster.
\
Twenty-six of 63 galaxies within 3 Abell radii are located within the central Abell radius. 
\
In this region, 5 objects were clearly classified by \citet{1973UGC...C...0000N} as early spiral galaxies, but only 3 early-type objects were identified in the region between 1 and 3 Abell radii.
\
Therefore early-type spiral galaxies, are predominantly found within the center of the cluster.    
\  
For late-type spiral galaxies, the ratio changes towards a deficiency in the core: 6 late-type galaxies were clearly identified within the central Abell radius, but 11 were identified between 1 and 3 Abell radii.
\
Five Sb/SBb galaxies were found in both the inner and the outer regions, thus resulting in a higher fraction for the center when referring to the total number of objects.
\   
Three galaxies were identified as peculiar, another 3 as dwarf galaxies. 
\
The remaining 22 galaxies were either identified as spirals without any further characterization or were not categorized at all.

However, for the L$_{\rm{FIR}}$ bright subsample and the CO subsample discussed in this paper, such a clear segregation is not visible.
\
Thirteen of the 36 objects with known FIR flux densities are located in the central Abell radius -- 4 late-type and 3 early-type galaxies -- whereas 7 late-type and 5 early type galaxies are found in the outer region between 1 and 3 Abell radii.
\
Of the  12 objects that have been observed in CO emission, 7 are located in the central region. Of these, 2 are early and 3 are late-type galaxies. One SBa and 2 late-type classifications were found for the outer objects. 
\
Neglecting the morphological distribution of the subsamples, therefore, should not result in a misinterpretation of L$_{\rm{FIR}}$ or CO-line intensity based quantities in terms of environmental influences.
\
\subsection{Molecular gas loss by stripping of a cirrus-like component?}
\label{SecCirrusStripping}
Gas clouds do not consist of atomic gas alone;  molecular gas and cold dust are also associated with them.
\
Several authors (e.g. \citealt{1986apj...311l..33h}; \citealt{1992mnras.258..787r}) have discussed the presence of cold dust in the ISM resulting in multiple-component models of galaxy emission in the IRAS wavelength domain.
\
Besides warmer emission from active star-forming regions and eventually a Seyfert component, a relatively constant disk or cirrus-like component from cold interstellar dust contributes to the IRAS colors of galaxies.

\begin{figure}
  \resizebox{\hsize}{!}{\includegraphics[]{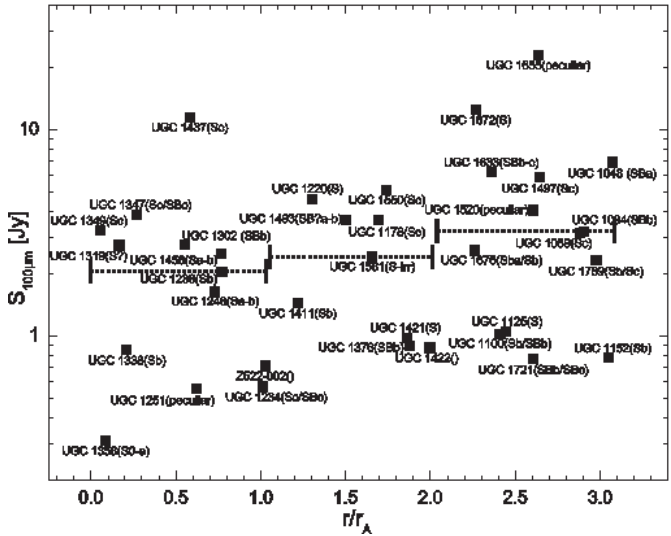}}
  \caption{IRAS 100$\mu$m flux densities of Abell~262 galaxies as a function of the distance to the cluster center. The plot contains all Abell~262 galaxies down to the IRAS flux limit with distances to the cluster core not exceeding 3 Abell radii. Median values of 3 regions are represented by the horizontal lines. }             
  \label{FigS100Distance}
\end{figure}

In the course of the HI removal processes, which these cirrus-like clouds in cluster galaxies are subjected to, it is likely that the associated molecular gas and cold dust are also lost.      
\
This is supported by \citet{1989mnras.239..347d}, who convincingly show the existence of dust deficiency in HI deficient galaxies in the core region of Virgo.  
\
By analyzing the FIR properties of the galaxies, they discuss the lack of dust in the context of the HI removal.
\
Since dust must be associated with molecular gas, there must be a CO deficiency present at a low level as well.
\
As shown in Sect. \ref{SecH2LossSignatures}, a FIR deficiency is indeed present 
in the set of IRAS detected galaxies in the core of Abell~262.
\
A total difference of  $\Delta {\rm S_{100\mu m}}\sim$ 1.1 Jy (Fig. \ref{FigS100Distance}) between the median values of the core (r$<$r$_{\rm A}$) and outer galaxies ($2{\rm r_{\rm A}<r<3r_{\rm A}}$) can be identified.

HI and FIR deficiency, as well as the other indications of a core distance dependency of the molecular gas content in Abell~262 galaxies, suggests the applicability of a model for cirrus-like cloud stripping.
\
This model is introduced in Sect. \ref{SecCirrusModel} and applied to the case of Abell~262 in Sect. \ref{SecApplyModel}.  
\
The derived estimates are not inconsistent with the presence of simultaneously occuring CO-deficiency in the core region.
\\
 
\begin{figure}
  \resizebox{8cm}{!}{\includegraphics[]{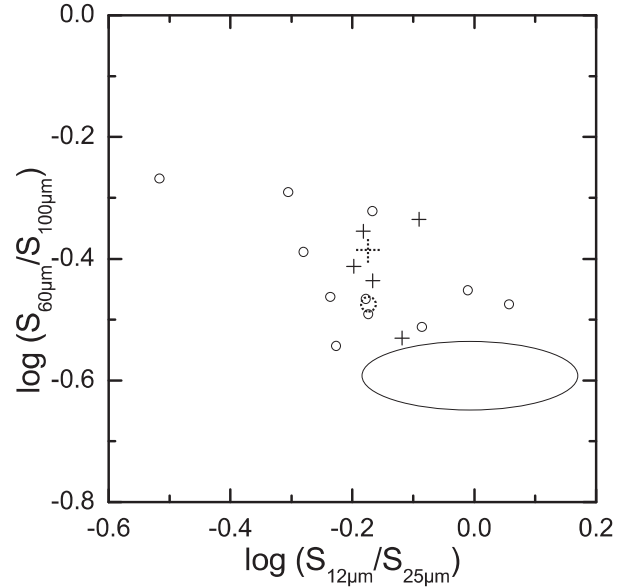}}
  \caption{Color-color diagram of Abell~262 galaxies detected in all four IRAS bands. Objects with a projected distance to the cluster core of less than 1 Abell radius are represented by a cross, and objects exceeding this distance are represented by open circles. The dotted symbols represent the respective median values. The ellipse marks the region of cirrus colors (cf. references in \citealt{1986apj...311l..33h}). The lowest cross belongs to UGC~1437, which is most likely a distant galaxy that is projected into the center. Therefore it was neglected in the median. In the median value of the more distant galaxies, only the objects with $\log{\rm S_{60\mu m}/S_{100\mu m}}<-0.4$ were considered. For the other objects a contribution of star formation to the FIR colors has to be assumed in addition to the cirrus-like component.}             
  \label{IRAS-Colors}
\end{figure}

Another indication that supports the stripping hypothesis of cirrus-like clouds can be found in the IRAS color-color diagram (cf. Fig. \ref{IRAS-Colors}). 
\
IRAS colors of normal galaxies are influenced by several factors. 
\
While the cirrus-like contribution is fairly constant for all galaxies, the importance of star formation varies and may significantly influence the galaxy's  position within the color-color diagram. 
\
The presence of  varying star-formation components on top of the cirrus-like component creates a distribution of galaxies within the color-color diagram that is non-Gaussian (cf. \citealt{1986apj...311l..33h}). 
\
Therefore, mean values are not adequate measures for comparing the colors of the cirrus-like components of central and more distant galaxies.
\
Objects with a strong star-formation component should be excluded from the analysis.
\
For the remaining objects, the median value of the core galaxies differs slightly from the value for the more distant galaxies, with the latter closer to that of typical cirrus colors.

\subsubsection{The galactic cirrus cloud model of \citet{1987apj...319..723d} }
\label{SecCirrusModel}
Galactic cirrus clouds, which are predominantly found at high galactic latitudes, consist of atomic and molecular gas.
\
Embedded in the gas is cold dust, which reradiates absorbed energy in the FIR.    
\
For the galactic cirrus in Ursa Major, \citet*{1987apj...319..723d} have proposed 
a two-component model to explain its infrared emission as a function of the total hydrogen column density, N(H)=N(HI)+2N(H$_2$): 
$${\rm I}_{\rm{100\mu m}}[{\rm MJy\,sr^{-1}}] = a {\rm N(HI)} + b \rm{I_{CO}}+I_{\rm{100\mu m,BG}}, $$        
where I$_{\rm{100\mu m,BG}}$ represents the flux density  of the background emission. Again $\rm{I_{CO}}$ is used as tracer for H$_2$: $X$=N(H$_2$)/I$_{\rm{CO}}$=$b/2a$.
\
Then \citet{1987apj...319..723d} derive a value for $a$ of  $(1.0\pm0.4)\cdot10^{-20}$MJy sr$^{-1}$ cm$^2$ and for $b$ of $(1.0\pm0.5)$ MJy sr$^{-1}$ K$^{-1}$ km$^{-1}$s.
\
In the cases of Centaurus~A \citep{1990apj...363..451e} and NGC 2903  \citep{1991apj...375..105j}, the scope of this model has already been sucessfully extended to extragalactic sources.
\subsubsection{Application of the model to the case of Abell~262}
\label{SecApplyModel}
Due to the large IRAS beam, S$_{\rm{100\mu m}}$ comprises the source-integrated flux density for each Abell~262 galaxy.
\
In order to extend the \citet{1987apj...319..723d} model to extragalactic sources, two assumptions are necessary: 1. the 100$\mu$m flux density traces N(H) on galactic scales, and 2. the background emission is similar over the whole extension of the cluster.
\
A difference between representative values of S$_{\rm{100\mu m}}$ for the cluster core and for the outer region may be proportional to:  
$$\Delta {\rm S_{100\mu m}} \propto {\textstyle  
  \left< \int_{\atop \rm Sou} N(H) {\rm d} \Omega \right>_{\rm out} - \left< \int_{\atop \rm Sou} N(H) {\rm d} \Omega \right>_{\rm core}.}$$
The source-integrated H column density splits into an HI and an H$_2$ component, which are directly related to the respective HI and H$_2$ masses. 
\
So does $\Delta {\rm S_{100\mu m}}$ in the model.
\
A direct determination of $\int N(H){\rm d} \Omega$ can be only made for the few galaxies with existing H$_2$ mass estimates,
but it is possible to obtain seperate estimates for the  $\Delta {\rm S_{100\mu m,HI}}$ and $\Delta {\rm S_{100\mu m,H_2}}$ components of $\Delta {\rm S_{100\mu m}}$, as we show in the following.    
\\  

The HI-deficiency HIDEF of galaxies in Abell~262, as defined and determined by \citet{1985apj...292..404g}, amounts to an average value of 0.33 for galaxies in the cluster's core region, with projected distances to the center of less than 1 Abell radius.
\
For galaxies in the outer region, with distances exceeding 1 Abell radius, the average value for HIDEF is 0. 
\
HIDEF refers to a comparison sample of isolated galaxies \citep{1984aj.....89..758h}, for which the authors claim the average mass of all galaxies to be log($h^2$M$_{\rm HI}$)=$9.39\pm0.52$.
\
HIDEF and the average mass of a galaxy in the comparison sample can be used to derive an average HI-mass difference between the core and the outer region, and hence a value for $\Delta {\rm S_{100\mu m,HI}}$ of 0.68$^{+1.58}_{-0.55}$Jy. 
\
The error mainly results from the uncertainties in the average HI mass and the factor $a$.   
\
The H$_2$ mass estimates given in Sect. \ref{SecMolecularMass} were used to obtain a gross estimate for $\Delta {\rm S_{100\mu m,H_2}}$ of 0.26$\pm$0.14 Jy.
\
Indeed, the estimates for the two components are in agreement with the 
the total difference  $\Delta {\rm S_{100\mu m}} \sim$ 1.1 Jy found in the IRAS data of Abell~262 galaxies. 

\section{Summary and conclusions}

Abell~262 is a spiral-rich cluster. 
\
Based on the classification given in Table \ref{Abell262-Properties}, no clear tendency toward morphological segregation can be identified in the samples discussed in this paper (cf. Sect. \ref{Morphology}).
\
The trend that the galaxies with less luminous CO-line emission and
FIR-emission are located towards the center of the cluster (i.e. within the Abell radius) suggests that these objects show a deficiency in molecular gas and dust.
\
Such an effect has not been reported so far. As in the case of the HI-deficiency, a CO-deficiency may be a result of interaction between neighboring galaxies or of interaction with either the central cluster potential or the central cluster gas reservoir.

The model  by \citet{1987apj...319..723d}, explaining the FIR emission of galactic cirrus clouds, can be generalized to the case of a global cirrus/disc component in galaxies.
\
One prediction of this model is a drop in the 100$\mu$m flux density ($\Delta {\rm S_{100\mu m}}$), which indeed is indicated in the existing IRAS data for the cluster. 
\
A small part of the decrease in $\Delta {\rm S_{100\mu m}}$ can be expected to be caused by a deficiency of molecular gas mass in the cluster, which is traced by I$_{\rm CO}$.
\
The derived contributions to $\Delta {\rm S_{100\mu m}}$ of 0.26$\pm$0.14 Jy for 
the molecular gas mass deficiency and of 0.68$^{+1.58}_{-0.55}$Jy for the atomic gas mass deficiency is consistent with the overall drop in the median 100 $\mu$m flux density of $\sim$1.1 Jy. 
\       
It can be expected that a certain amount of molecular gas is associated with the
more diffuse HI material.
\
This is especially the case for high-latitude cirrus clouds that are also bright in their FIR emission.
\
The weak indications of a CO- and L$_{\rm FIR}$-deficiency observed in Abell~262 may, therefore, be a direct consequence of the HI-deficiency found by \citet*{1982apj...262..442g}.
\\

As for UGC~1347, a second bright source south of the nucleus can be identified not only in the visual and NIR but also in the mm wavelength domain. 
\
Both sources are optically thick in their CO line emmission.
\
The velocity profile is in good agreement with a previously published H$\alpha$ rotation curve. 
\
Within the position-velocity diagram, the southern component deviates from the rotation curve of the disk. 
\ 
This may indicate an asymmetrically expanding gas shell or a large scale outflow at the position of the southern source, which supports a scenario of enhaced star formation in that region, as was indicated by \citet{2000a&a...363...41h}.   
 
\begin{acknowledgements} 
This work was supported in part by the Deutsche Forschungsgemeinschaft (DFG) via grant SFB 494.
\end{acknowledgements}
 
\bibliographystyle{aa}
\bibliography{2564bib}

\begin{thebibliography}{46}
\expandafter\ifx\csname natexlab\endcsname\relax\def\natexlab#1{#1}\fi

\bibitem[{{Abell}(1958)}]{1958apjs....3..211a}
{Abell}, G.~O. 1958, \apjs, 3, 211

\bibitem[{{Abell} {et~al.}(1989){Abell}, {Corwin}, \&
  {Olowin}}]{1989apjs...70....1a}
{Abell}, G.~O., {Corwin}, H.~G., \& {Olowin}, R.~P. 1989, \apjs, 70, 1

\bibitem[{{Amram} {et~al.}(1994){Amram}, {Marcelin}, {Balkowski}, {Cayatte},
  {Sullivan}, \& {Le Coarer}}]{1994a&as..103....5a}
{Amram}, P., {Marcelin}, M., {Balkowski}, C., {et~al.} 1994, \aaps, 103, 5

\bibitem[{{Boselli} {et~al.}(1997){Boselli}, {Gavazzi}, {Lequeux}, {Buat},
  {Casoli}, {Dickey}, \& {Donas}}]{1997a&a...327..522b}
{Boselli}, A., {Gavazzi}, G., {Lequeux}, J., {et~al.} 1997, \aap, 327, 522

\bibitem[{{Bravo-Alfaro} {et~al.}(1997){Bravo-Alfaro}, {Szomoru}, {Cayatte},
  {Balkowski}, \& {Sancisi}}]{1997A&AS..126..537B}
{Bravo-Alfaro}, H., {Szomoru}, A., {Cayatte}, V., {Balkowski}, C., \&
  {Sancisi}, R. 1997, \aaps, 126, 537

\bibitem[{{Butcher} \& {Oemler}(1978)}]{1978apj...226..559b}
{Butcher}, H. \& {Oemler}, A. 1978, \apj, 226, 559

\bibitem[{{Casoli} {et~al.}(1991){Casoli}, {Boisse}, {Combes}, \&
  {Dupraz}}]{1991a&a...249..359c}
{Casoli}, F., {Boisse}, P., {Combes}, F., \& {Dupraz}, C. 1991, \aap, 249, 359

\bibitem[{{Casoli} {et~al.}(1998){Casoli}, {Sauty}, {Gerin}, {Boselli},
  {Fouque}, {Braine}, {Gavazzi}, {Lequeux}, \& {Dickey}}]{1998a&a...331..451c}
{Casoli}, F., {Sauty}, S., {Gerin}, M., {et~al.} 1998, \aap, 331, 451

\bibitem[{{Couch} {et~al.}(1998){Couch}, {Barger}, {Smail}, {Ellis}, \&
  {Sharples}}]{1998apj...497..188c}
{Couch}, W.~J., {Barger}, A.~J., {Smail}, I., {Ellis}, R.~S., \& {Sharples},
  R.~M. 1998, \apj, 497, 188

\bibitem[{{David} {et~al.}(1996){David}, {Jones}, \&
  {Forman}}]{1996ApJ...473..692D}
{David}, L.~P., {Jones}, C., \& {Forman}, W. 1996, \apj, 473, 692

\bibitem[{{de Vaucouleurs} {et~al.}(1991){de Vaucouleurs}, {de Vaucouleurs},
  {Corwin}, {Buta}, {Paturel}, \& {Fouque}}]{1991trcb.book.....D}
{de Vaucouleurs}, G., {de Vaucouleurs}, A., {Corwin}, H.~G., {et~al.} 1991,
  {Third Reference Catalogue of Bright Galaxies} (Springer-Verlag Berlin
  Heidelberg New York)

\bibitem[{{de Vries} {et~al.}(1987){de Vries}, {Thaddeus}, \&
  {Heithausen}}]{1987apj...319..723d}
{de Vries}, H.~W., {Thaddeus}, P., \& {Heithausen}, A. 1987, \apj, 319, 723

\bibitem[{{Doyon} \& {Joseph}(1989)}]{1989mnras.239..347d}
{Doyon}, R. \& {Joseph}, R.~D. 1989, \mnras, 239, 347

\bibitem[{{Eckart} {et~al.}(1990{\natexlab{a}}){Eckart}, {Cameron},
  {Rothermel}, {Wild}, {Zinnecker}, {Rydbeck}, {Olberg}, \&
  {Wiklind}}]{1990apj...363..451e}
{Eckart}, A., {Cameron}, M., {Rothermel}, H., {et~al.} 1990{\natexlab{a}},
  \apj, 363, 451

\bibitem[{{Eckart} {et~al.}(1990{\natexlab{b}}){Eckart}, {Downes}, {Genzel},
  {Harris}, {Jaffe}, \& {Wild}}]{1990ApJ...348..434E}
{Eckart}, A., {Downes}, D., {Genzel}, R., {et~al.} 1990{\natexlab{b}}, \apj,
  348, 434

\bibitem[{{Fadda} {et~al.}(1996){Fadda}, {Girardi}, {Giuricin}, {Mardirossian},
  \& {Mezzetti}}]{1996ApJ...473..670F}
{Fadda}, D., {Girardi}, M., {Giuricin}, G., {Mardirossian}, F., \& {Mezzetti},
  M. 1996, \apj, 473, 670

\bibitem[{{Giovanelli} \& {Haynes}(1985)}]{1985apj...292..404g}
{Giovanelli}, R. \& {Haynes}, M.~P. 1985, \apj, 292, 404

\bibitem[{{Giovanelli} {et~al.}(1982){Giovanelli}, {Haynes}, \&
  {Chincarini}}]{1982apj...262..442g}
{Giovanelli}, R., {Haynes}, M.~P., \& {Chincarini}, G.~L. 1982, \apj, 262, 442

\bibitem[{{Hackenberg} {et~al.}(2000){Hackenberg}, {Eckart}, {Davies},
  {Rabien}, {Ott}, {Kasper}, {Hippler}, \& {Quirrenbach}}]{2000a&a...363...41h}
{Hackenberg}, W., {Eckart}, A., {Davies}, R.~I., {et~al.} 2000, \aap, 363, 41

\bibitem[{{Haynes} \& {Giovanelli}(1984)}]{1984aj.....89..758h}
{Haynes}, M.~P. \& {Giovanelli}, R. 1984, \aj, 89, 758

\bibitem[{{Haynes} \& {Giovanelli}(1986)}]{1986apj...306l..55h}
{Haynes}, M.~P. \& {Giovanelli}, R. 1986, \apjl, 306, L55

\bibitem[{{Haynes} {et~al.}(1997){Haynes}, {Giovanelli}, {Herter}, {Vogt},
  {Freudling}, {Maia}, {Salzer}, \& {Wegner}}]{1997AJ....113.1197H}
{Haynes}, M.~P., {Giovanelli}, R., {Herter}, T., {et~al.} 1997, \aj, 113, 1197

\bibitem[{{Helfer} {et~al.}(2002){Helfer}, {Vogel}, {Lugten}, \&
  {Teuben}}]{2002pasp..114..350h}
{Helfer}, T.~T., {Vogel}, S.~N., {Lugten}, J.~B., \& {Teuben}, P.~J. 2002,
  \pasp, 114, 350

\bibitem[{{Helou}(1986)}]{1986apj...311l..33h}
{Helou}, G. 1986, \apjl, 311, L33

\bibitem[{{Helou} {et~al.}(1988){Helou}, {Khan}, {Malek}, \&
  {Boehmer}}]{1988apjs...68..151h}
{Helou}, G., {Khan}, I.~R., {Malek}, L., \& {Boehmer}, L. 1988, \apjs, 68, 151

\bibitem[{{Horellou} {et~al.}(1995){Horellou}, {Casoli}, \&
  {Dupraz}}]{1995A&A...303..361H}
{Horellou}, C., {Casoli}, F., \& {Dupraz}, C. 1995, \aap, 303, 361

\bibitem[{{Israel}(2001)}]{2001mhs..conf..293I}
{Israel}, F. 2001, in F. Combes, and G. Pineau des For�ts (eds), Molecular
  hydrogen in space, Cambridge University Press, Cambridge, UK, 293

\bibitem[{{Israel}(1997)}]{1997A&A...328..471I}
{Israel}, F.~P. 1997, \aap, 328, 471

\bibitem[{{Jackson} {et~al.}(1991){Jackson}, {Eckart}, {Cameron}, {Wild}, {Ho},
  {Pogge}, \& {Harris}}]{1991apj...375..105j}
{Jackson}, J.~M., {Eckart}, A., {Cameron}, M., {et~al.} 1991, \apj, 375, 105

\bibitem[{{Jones} \& {Forman}(1984)}]{1984ApJ...276...38J}
{Jones}, C. \& {Forman}, W. 1984, \apj, 276, 38

\bibitem[{{Kenney} \& {Young}(1989)}]{1989apj...344..171k}
{Kenney}, J.~D.~P. \& {Young}, J.~S. 1989, \apj, 344, 171

\bibitem[{{Lavezzi} \& {Dickey}(1998)}]{1998aj....115..405l}
{Lavezzi}, T.~E. \& {Dickey}, J.~M. 1998, \aj, 115, 405

\bibitem[{{MacKenzie} {et~al.}(1996){MacKenzie}, {Schlegel}, \&
  {Mushotzky}}]{1996ApJ...468...86M}
{MacKenzie}, M., {Schlegel}, E.~M., \& {Mushotzky}, R. 1996, \apj, 468, 86

\bibitem[{{Moshir} {et~al.}(1990){Moshir}, {Copan}, {Conrow}, {McCallon},
  {Hacking}, {Gregorich}, {Rohrbach}, {Melnyk}, {Rice}, {Fullmer}, \&
  {Chester}}]{1990irasf.c......0m}
{Moshir}, M., {Copan}, G., {Conrow}, T., {et~al.} 1990, {IRAS Faint Source
  Catalogue, Version 2.0.}

\bibitem[{{Moss} \& {Whittle}(2000)}]{2000mnras.317..667m}
{Moss}, C. \& {Whittle}, M. 2000, \mnras, 317, 667

\bibitem[{{Neill} {et~al.}(2001){Neill}, {Brodie}, {Craig}, {Hailey}, \&
  {Misch}}]{2001ApJ...548..550N}
{Neill}, J.~D., {Brodie}, J.~P., {Craig}, W.~W., {Hailey}, C.~J., \& {Misch},
  A.~A. 2001, \apj, 548, 550

\bibitem[{{Nilson}(1973)}]{1973UGC...C...0000N}
{Nilson}, P. 1973, {Uppsala General Catalogue of Galaxies, Acta Universitatis
  Upsalienis, Nova Acta Regiae Societatis Upsaliensis, Series V: A, Vol. 1}

\bibitem[{{Rowan-Robinson}(1992)}]{1992mnras.258..787r}
{Rowan-Robinson}, M. 1992, \mnras, 258, 787

\bibitem[{{Sanders} {et~al.}(1991){Sanders}, {Scoville}, \&
  {Soifer}}]{1991ApJ...370..158S}
{Sanders}, D.~B., {Scoville}, N.~Z., \& {Soifer}, B.~T. 1991, \apj, 370, 158

\bibitem[{{Solanes} {et~al.}(2001){Solanes}, {Manrique}, {Garc{\'{\i}}a-G{\'
  o}mez}, {Gonz{\' a}lez-Casado}, {Giovanelli}, \&
  {Haynes}}]{2001apj...548...97s}
{Solanes}, J.~M., {Manrique}, A., {Garc{\'{\i}}a-G{\' o}mez}, C., {et~al.}
  2001, \apj, 548, 97

\bibitem[{{Solomon} \& {Sage}(1988)}]{1988apj...334..613s}
{Solomon}, P.~M. \& {Sage}, L.~J. 1988, \apj, 334, 613

\bibitem[{{Stark} {et~al.}(1986){Stark}, {Knapp}, {Bally}, {Wilson}, {Penzias},
  \& {Rowe}}]{1986apj...310..660s}
{Stark}, A.~A., {Knapp}, G.~R., {Bally}, J., {et~al.} 1986, \apj, 310, 660

\bibitem[{{Strong} {et~al.}(1988){Strong}, {Bloemen}, {Dame}, {Grenier},
  {Hermsen}, {Lebrun}, {Nyman}, {Pollock}, \& {Thaddeus}}]{1988A&A...207....1S}
{Strong}, A.~W., {Bloemen}, J.~B.~G.~M., {Dame}, T.~M., {et~al.} 1988, \aap,
  207, 1

\bibitem[{{Struble} \& {Rood}(1999)}]{1999apjs..125...35s}
{Struble}, M.~F. \& {Rood}, H.~J. 1999, \apjs, 125, 35

\bibitem[{{Vollmer} {et~al.}(2001){Vollmer}, {Braine}, {Balkowski}, {Cayatte},
  \& {Duschl}}]{2001a&a...374..824v}
{Vollmer}, B., {Braine}, J., {Balkowski}, C., {Cayatte}, V., \& {Duschl}, W.~J.
  2001, \aap, 374, 824

\bibitem[{{Young} {et~al.}(1995){Young}, {Xie}, {Tacconi}, {Knezek}, {Viscuso},
  {Tacconi-Garman}, {Scoville}, {Schneider}, {Schloerb}, {Lord}, {Lesser},
  {Kenney}, {Huang}, {Devereux}, {Claussen}, {Case}, {Carpenter}, {Berry}, \&
  {Allen}}]{1995ApJS...98..219Y}
{Young}, J.~S., {Xie}, S., {Tacconi}, L., {et~al.} 1995, \apjs, 98, 219

\end{thebibliography}

\end{document}